\documentclass[twocolumn,aps,prd,preprintnumbers,groupedaddress,tightenlines,nofootinbib,floatfix,10pt,reprint,superscriptaddress]{revtex4-2}

\usepackage{amssymb,amsmath,bm,graphicx,hyperref,xcolor}

\hypersetup{
    colorlinks=true,       
    linkcolor=blue,          
    citecolor=blue,        
    urlcolor=blue,           
}

\begin{document}

\preprint{J-PARC-TH-0246,UTHEP-758}

\title{Finite-size scaling around the critical point in the heavy quark region of QCD
}

\author{Atsushi Kiyohara}
\affiliation{
  Department of Physics, Osaka University, Toyonaka, Osaka 560-0043, Japan}
\author{Masakiyo Kitazawa}
\affiliation{
  Department of Physics, Osaka University, Toyonaka, Osaka 560-0043, Japan}
\affiliation{
  J-PARC Branch, KEK Theory Center, 
  Institute of Particle and Nuclear Studies, KEK, 
  203-1, Shirakata, Tokai, Ibaraki 319-1106, Japan}
\author{Shinji Ejiri}
\affiliation{
  Department of Physics, Niigata University, Niigata 950-2181, Japan}
\author{Kazuyuki Kanaya}
\affiliation{
  Tomonaga Center for the History of the Universe, University of Tsukuba, Tsukuba, Ibaraki 305-8571, Japan}

\date{\today}

\begin{abstract}

  Finite-size scaling is investigated in detail around the critical point
  in the heavy-quark region of nonzero temperature QCD. 
  Numerical simulations are performed with large spatial volumes
  up to the aspect ratio $N_s/N_t=12$ at a fixed lattice spacing with $N_t=4$.
  We show that the Binder cumulant and
  the distribution function of the Polyakov loop follow
  the finite-size scaling in the $Z(2)$ universality class for
  large spatial volumes with $N_s/N_t \ge 9$, while, for $N_s/N_t \le 8$, the Binder cumulant becomes inconsistent with the $Z(2)$ scaling.
  To realize the large-volume simulations in the heavy-quark region, 
  we adopt the hopping parameter expansion for the quark determinant:
  We generate gauge configurations using the leading order action including the Polyakov loop term for $N_t=4$,
  and incorporate the next-to-leading order effects in the measurements by the multipoint reweighting method.
  We find that the use of the leading-order configurations
  is crucially effective in suppressing the overlapping problem in the reweighting and thus reducing the statistical errors.  
  
\end{abstract}

\maketitle

\section{Introduction}
\label{sec:intro}

One of the interesting features of the medium described by 
quantum chromodynamics (QCD) is the existence of phase transitions
of various orders.
While the finite-temperature QCD transition is an analytic crossover at zero
quark chemical potential $\mu_q$~\cite{Aoki:2006we,Ejiri:2009ac},
this phase transition
is expected to become of first order in dense
medium with large $\mu_q$~\cite{Asakawa:1989bq}.
The end point of the first-order transition is called 
the critical point (CP) at which the transition is of second order.
The singularity in thermodynamic observables associated with the second order CP are believed to be useful in detecting the CP in 
heavy-ion collision experiments~\cite{Stephanov:1999zu,Asakawa:2015ybt}.
Accordingly, researches called
the beam-energy scan are actively performed in experimental facilities
all over the world to search for the critical fluctuations
around the CP~\cite{Asakawa:2015ybt,Bzdak:2019pkr,Bluhm:2020mpc}.

The order of the finite temperature QCD transition 
changes also with variation of quark masses~\cite{Pisarski:1983ms}.
For the $2+1$-flavor QCD, it is known that the crossover 
at the physical quark masses 
becomes of first order both in the light- and heavy-quark limits;
the phase diagram representing this feature is known as the Columbia
plot~\cite{Brown:1990ev,Gavin:1993yk}. 
Revealing the nature of the phase transitions with the variation of quark
masses is an important subject of QCD at nonzero temperature 
since it provides us with various insights into the transition
at the physical quark masses.

Pinning down the boundaries of the first-order transitions
in $2+1$-flavor QCD in the 
light~\cite{Iwasaki:1995yj,Iwasaki:1996zt,JLQCD:1998mja,Karsch:2001nf,deForcrand:2003vyj,Bernard:2004je,deForcrand:2006pv,Cheng:2006aj,Jin:2013wta,Ejiri:2015vip,Bazavov:2017xul,Jin:2017jjp,Kuramashi:2020meg} and
heavy~\cite{Banks:1983me,DeGrand:1983fk,Alexandrou:1998wv,Saito:2011fs,Fromm:2011qi,Saito:2013vja,Ejiri:2019csa,Cuteri:2020yke}
quark regions is a longstanding subject in lattice QCD simulations.
It, however, has been found that the location of the boundaries 
are strongly dependent on the lattice cutoff of the
simulations~\cite{Jin:2017jjp,Kuramashi:2020meg,Cuteri:2020yke},
and their quantitative determination in the continuum limit has not been established yet.

One of the difficulties in these analyses 
is that observables near the CP
are strongly dependent on the spatial volume of the system.
The spatial volume dependence is in part described by the
finite-size scaling (FSS)~\cite{Pelissetto:2000ek}.
However, the FSS is applicable only for describing the singular part of
thermodynamic quantities that dominates over the non-singular part
only in the vicinity of the CP for sufficiently large spatial volumes.
When the spatial volume is not large enough the FSS of observables
is violated due to the contributions of the non-singular part and 
this makes their analysis based on the FSS problematic.
In fact, although the CP of QCD is believed to belong to the
three-dimensional $Z(2)$ universality class~\cite{Pisarski:1983ms}, 
a clear FSS in this universality class has not been observed
in the latest numerical study around the CP 
in the light quark mass region~\cite{Kuramashi:2020meg}.
In the heavy quark region, 
on lattices of $N_t=6$ and 8 with the aspect ratio $N_s/N_t=6$ and 8, 
the Binder cumulant $B_4$ 
is reported to be consistent with the $Z(2)$ FSS using data mostly in the crossover side, 
while data in the first order side as well as that on $N_t=10$ lattice show deviation from the $Z(2)$ FSS~\cite{Cuteri:2020yke}. 
These results suggest the necessity to perform numerical analyses
with yet larger spatial volumes and with high statistics.

In the present study, we focus on the CP in the heavy quark region
and study the behavior of observables by numerical simulations
with large spatial volumes corresponding to the aspect ratios up to $N_s/N_t=12$.
To carry out analyses on the large spatial volumes with high precision, 
we fix the temporal lattice extent to be $N_t=4$ in this study.

We also employ the hopping parameter expansion (HPE) to deal with 
the quark determinant.
In this study, we generate gauge configurations using
the leading order (LO) action of the HPE including the Polyakov loop term for $N_t=4$,
and then incorporate the next-to-leading order (NLO) effects by a multipoint reweighting method~\cite{Ferrenberg:1989ui}. 
In our previous study, we generated the configurations in quenched QCD (zero-th order of the HPE) and reweighted them to incorporate LO and NLO effects~\cite{Saito:2011fs,Saito:2013vja,Ejiri:2019csa}.
We show that the use of the LO action to generate configurations is quite effective in
suppressing the overlapping problem of the reweighting method. 
This is essential to carry out simulations with large system volumes as studied in the present paper.
We also verify the convergence of the HPE by comparing
the LO and NLO results.

We perform the Binder cumulant analysis of the Polyakov loop
around the CP and find that the numerical results for $N_s/N_t\ge9$ follow well 
the FSS in the $Z(2)$ universality class.
On the other hand, inclusion of the data at $N_s/N_t\le8$ in the analysis
gives rise to a statistically-significant deviation from the scaling behavior.
We further investigate the scaling behavior of the distribution function
of the Polyakov loop.
We find that the distribution function follows the FSS 
in the $Z(2)$ universality class for 
$N_s/N_t\ge8$.
From the deviation pattern of the distribution function for $LT=6$ from the $Z(2)$ FSS, we discuss that the violation of the scaling behavior in the Binder cumulant
is caused by the deviation in the tails of the distribution for small $LT$. 
In this paper, we consider the case of degenerated $N_{\rm f}$ flavors
with $N_{\rm f}=1,2,3$.
Generalization of the formalism to non-degenerate cases is straightforward.

This paper is organized as follows.
In the next section we give a brief review on the FSS.
We then explain the setup of our lattice simulation and analyses using the HPE in Sec.~\ref{sec:setup}.
In Sec.~\ref{sec:Binder},
we determine the transition line and perform the Binder cumulant analysis to determine the
location of the CP and the critical exponent.
In Sec.~\ref{sec:p(O)},
we investigate the FSS of the distribution function of the Polyakov loop.
The last section, Sec.~\ref{sec:conclusion}, is devoted to a summary.
In Appendix~\ref{sec:cumulant}, we give definition of cumulants.
In Appendix~\ref{sec:smear}, we examine the effect of the smearing width used in the calculation of distribution functions.
In Appendix~\ref{sec:HPEapp}, the HPE of the
quark determinant is calculated up to the NLO.
In Appendix~\ref{sec:LO}, the convergence of the HPE is examined by comparing the Binder cumulants at the LO and the NLO.

\section{Finite-size scaling}
\label{sec:FSS}

\begin{figure}
  \centering
    \includegraphics[width=0.48\textwidth]{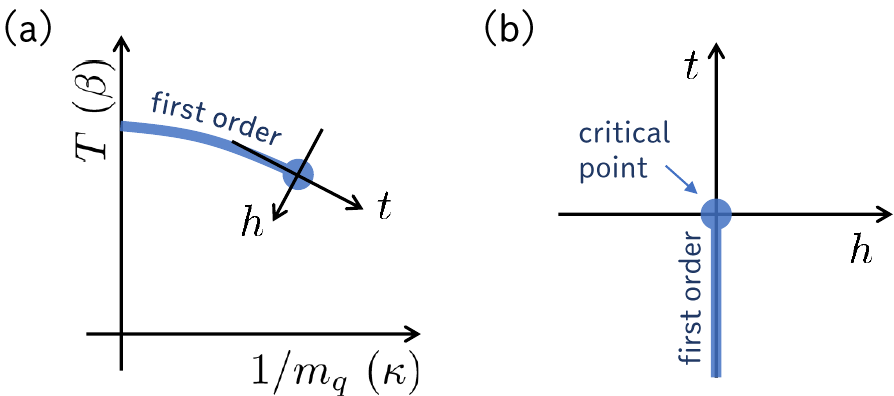}
  \caption{
    Phase diagrams of (a) QCD in heavy-quark region and (b) three-dimensional Ising model.
  }
\label{fig:scaling}
\end{figure}

Let us first give a brief review on the FSS and its application to
the CP in the heavy-quark region of QCD.

The heavy-quark limit of QCD corresponds to the $SU(3)$ Yang-Mills theory (quenched QCD).
This theory has a first-order deconfinement phase transition
at nonzero temperature $T$.
When the quark mass $m_q$ is finite (throughout this section we assume that the quark masses are degenerate), with increasing $1/m_q$,
this first-order transition becomes weaker and eventually
terminates at the CP, 
as schematically shown in the phase diagram on the $(T,1/m_q)$
plane in Fig.~\ref{fig:scaling}~(a).
This CP, as well as that in the light quark region, 
is believed to belong to the $Z(2)$ universality class, \textit{i.e.}
the universality class of the three-dimensional Ising
model~\cite{Pisarski:1983ms}.

Near the CP of the three-dimensional Ising model, the relevant scaling
parameters are the reduced temperature $t$ and external magnetic field $h$;
extensive variables conjugate to these parameters are the energy and the 
magnetization, respectively.
As shown in Fig.~\ref{fig:scaling}~(b), the CP 
is located at $(t,h)=(0,0)$ and the first-order transition
exists on the $t$ axis for $t<0$.
The singular part of thermodynamic quantities near the CP is
described by the scaling function of $t$ and $h$.
According to the universality, the singular part of thermodynamic quantities
near the CP of heavy-quark QCD is described by the
same scaling function, where 
the scaling parameters $t$ and $h$ are encoded into the $(T,1/m_q)$ plane
as schematically shown in Fig.~\ref{fig:scaling}~(a); the $t$ axis is
parallel to the first-order line at the CP while the direction of the
$h$ axis is not constrained from the universality.

According to the FSS argument~\cite{Pelissetto:2000ek}
the singular part of the dimension-less free energy $F(t,h,L^{-1})$ around the CP 
obtained at a finite volume $V=L^3$ has a scaling 
\begin{align}
  F(t,h,L^{-1})
  = F(tb^{y_t} , hb^{y_h} , L^{-1}b ),
  \label{eq:F}
\end{align}
for arbitrary scale factor $b$.
The values of the exponents
$y_t$ and $y_h$ are specific for each universality class.
In the $Z(2)$ universality class these parameters
are numerically obtained as~\cite{Pelissetto:2000ek}
\begin{align}
  y_t= 1.588, \quad
  y_h= 2.482.
  \label{eq:Z2y}
\end{align}
By setting $b=L$ one has
\begin{align}
  F(t,h,L^{-1})
  = F(tL^{y_t} , hL^{y_h} , 1 )
  \equiv \tilde{F}(tL^{y_t} , hL^{y_h} ).
  \label{eq:Ftilde}
\end{align}

Derivatives of $F(t,h,L^{-1})$ with respect to $t$ and $h$ define
the cumulants of the corresponding extensive variables.
For example, cumulants of the magnetization $M$ are given
by
\begin{align}
  \langle M(t,h,L^{-1})^n \rangle_{\rm c}
  = \partial_h^n F(t,h,L^{-1}),
  \label{eq:<M^n>c}
\end{align}
with $\partial_h=\partial/\partial h$.
From Eq.~(\ref{eq:Ftilde}), $L$ dependence of the cumulants
$\langle M^n \rangle_{\rm c}$ near the CP are written as 
\begin{align}
  \langle M(t,h,L^{-1})^n  \rangle_{\rm c}
  &= \partial_h^n F(t,h,L^{-1})
  \nonumber \\
  &= L^{ny_h} \partial_h^n \tilde{F}(tL^{y_t} , hL^{y_h} ).
  \label{eq:<M^n>cL}
\end{align}

In Ref.~\cite{Binder:1981sa}, it is suggested that the 
so-called (fourth-order) Binder cumulant 
\begin{align}
  B_4(t,h,L^{-1})
  &= \frac{ \langle M(t,h,L^{-1})^4 \rangle_{\rm c} }
  { (\langle M(t,h,L^{-1})^2 \rangle_{\rm c})^2 } + 3
  \label{eq:B4}
\end{align}
plays a useful role to determine the location of
the CP from numerical results obtained at finite $L$.
When the distribution of $M$ obeys
the Gauss distribution or the distribution composed of two
delta functions with an equal weight,
we have
\begin{align}
  B_4 =
  \begin{cases}
    3 & \mathrm{Gauss~distribution} , \\
    1 & \mathrm{two~delta~functions},
  \end{cases}
  \label{eq:B4lim}
\end{align}
respectively.
Since the distribution of $M$ approaches these functions
in the $L\to\infty$ limit on the crossover and first-order lines at $h=0$, respectively,
$B_4$ should approach Eq.~(\ref{eq:B4lim}) on these lines in the $L\to\infty$ limit.
Moreover, from Eq.~(\ref{eq:<M^n>c}), $B_4$ at $h=0$ behaves
as a function of $t$ and $L$ as 
\begin{align}
  B_4(t,0,L^{-1})
  &= \frac{ \partial_h^4 F(t , 0 , L^{-1}) }
  { (\partial_h^2 F(t , 0 , L^{-1} ))^2 } + 3
  \nonumber \\
  &= \frac{ \partial_h^4 \tilde{F}(tL^{y_t} , 0 ) }
            { (\partial_h^2 \tilde{F}(tL^{y_t} , 0 ))^2 } + 3
  \nonumber \\
  &= b_4 + c tL^{1/\nu} + {\cal O}(t^2),
  \label{eq:B4L}
\end{align}
for small $t$, where $b_4 = \partial_h^4 \tilde{F}(0,0) / (\partial_h^2 \tilde{F}(0,0 ))^2 + 3 $, 
$\nu=1/y_t$, and $c$ is a constant.
Eq.~(\ref{eq:B4L}) shows that $B_4(t,0,L^{-1})$ obtained for various $L$
at $h=0$ has a crossing at $t=0$.
The parameter $b_4$ is given only from $\tilde{F}(t,h)$
and thus are specific for each universality class.
For the $Z(2)$ universality class,
the value is known to be~\cite{Pelissetto:2000ek}
\begin{align}
  b_4  = 1.604 .
  \label{eq:Z2b4}
\end{align}

Equation~(\ref{eq:<M^n>c}) means that 
$F(t,h,L^{-1})$ is the cumulant generating function of $M$
up to an additive constant. 
Then, as shown in Eqs.~(\ref{eq:G(theta)}) and (\ref{eq:K})
in Appendix~\ref{sec:cumulant}, 
this function is related to the probability distribution function $p_M(M;t,h,L^{-1})$
of $M$ as
\begin{align}
  e^{F(t,h'-h,L^{-1})} = c_F \int dM \, e^{h'M} p_M(M;t,h,L^{-1}),
  \label{eq:e^F}
\end{align}
where $c_F$ is a constant determined from the normalization
condition $\int dM \, p_M(M)=1$.
Here, let us define another probability distribution
$\tilde{p}_M(M;t)$ as
\begin{align}
  e^{\tilde{F}(t,h')} = c_{\tilde{F}} \int dM e^{h'm} \tilde{p}_M(M;t).
  \label{eq:e^Ftilde}
\end{align}
From Eq.~(\ref{eq:Ftilde}), one finds 
\begin{align}
  p_M(M;t,0,L^{-1}) = L^{y_h} \tilde{p}_M(ML^{-y_h};tL^{y_t}).
\end{align}
When we consider magnetization
per unit volume $m=M/V$, the probability distribution of $m$ is given by 
\begin{align}
  p_m(m;t,0,L^{-1}) = L^{y_h-3} \tilde{p}_M(mL^{3-y_h};tL^{y_t}).
  \label{eq:p_m}
\end{align}
At the CP, $(t,h)=(0,0)$, one finds from Eq.~(\ref{eq:p_m}) that
\begin{align}
  p_m(m;0,0,L^{-1}) = L^{y_h-3} \tilde{p}_M( mL^{3-y_h} ; 0 ).
  \label{eq:p_m-CP}
\end{align}
Equation~(\ref{eq:p_m}) also suggests that, 
when $\tilde{p}_M(M;t)$ has a local extremum at $M=\tilde{M}(t)$,
$p_m(m;t,0,L^{-1})$ has corresponding local extremum at 
\begin{align}
  m = L^{y_h-3} \tilde{M}(tL^{y_t} ).
  \label{eq:minimum}
\end{align}
This implies that the $t$ and $L$ dependences of the
maximum of $p_m(m;t,0,L^{-1})$ are described by 
a single function $\tilde{M}(t)$.

\section{Setup}
\label{sec:setup}

\subsection{Lattice action and basic observables}
\label{sec:action}

In this study we investigate the four-dimensional system 
described by the lattice action of QCD
\begin{align}
S = S_g + S_q,
\label{eq:Stot}
\end{align} 
with $S_g$ and $S_q$ being the gauge and quark actions.
For $S_g$ we employ the plaquette action 
\begin{align}
S_g = -6 N_{\rm site} \,\beta \, \hat{P},
\label{eq:Sg}
\end{align} 
with the gauge coupling parameter $\beta = 6/g^2$ and
the space-time lattice volume $N_{\rm site} = N_s^3 \times N_t$.
The plaquette operator $\hat{P}$ is given by
\begin{align}
\hat{P}= \frac{1}{6 N_{\rm c} N_{\rm site}} \displaystyle \sum_{x,\,\mu < \nu} 
 {\rm Re \ tr_C} \left[ U_{x,\mu} U_{x+\hat{\mu},\nu}
U^{\dagger}_{x+\hat{\nu},\mu} U^{\dagger}_{x,\nu} \right] ,
\label{eq:Plaq}
\end{align} 
where $U_{x,\mu}$ is the link variable in the $\mu$ direction at site $x$, 
$x+ \hat\mu$ is the next site in the $\mu$ direction from $x$, $N_{\rm c}=3$, 
and ${\rm tr_C}$ is the trace over color indices.

For $S_q$, we adopt the Wilson quark action 
\begin{align}
  S_q = \sum_{f=1}^{N_{\rm f}} \sum_{x,\,y} \bar{\psi}_x^{(f)} \,
  M_{xy} (\kappa_f) \, \psi_y^{(f)} ,
\label{eq:Sq}
\end{align} 
with the Wilson quark kernel
\begin{align}
  &M_{xy} (\kappa_f) = \delta_{xy} - \kappa_f B_{xy},
  \label{eq:Mxy}
  \\
  &B_{xy}
  =  \sum_{\mu=1}^4 \left[ (1-\gamma_{\mu})\,U_{x,\mu}\,\delta_{y,x+\hat{\mu}} + (1+\gamma_{\mu})\,U_{y,\mu}^{\dagger}\,\delta_{y,x-\hat{\mu}} \right],
  \label{eq:B}
\end{align} 
where $x$, $y$ represent lattice sites.
The color and Dirac-spinor indices are suppressed for simplicity.
$\kappa_f$ is the hopping parameter for the $f$th flavor.
The bare quark mass $m_f$ is related to $\kappa_f$ as
\begin{align}
  \kappa_f = \frac1{2am_f+8} ,
\end{align}
with the lattice spacing $a$.
The matrix $B_{xy}$ has nonzero values only when 
lattice sites $x$ and $y$ are located in an adjacent sites.
Therefore, this term represents the ``hopping'' of
a quark between adjacent sites.
The heavy quark limit $m_f\to\infty$ corresponds to $\kappa_f\to0$.

In the following, we consider degenerated $N_{\rm f}$ flavors
with a common hopping parameter $\kappa=\kappa_f$
corresponding to a common quark mass $m_q$ --- generalization to non-degenerate cases is straightforward.
In this case, the expectation value of a gauge operator $\hat{O}(U)$
is calculated as
\begin{align}
  \langle \hat{O}(U) \rangle
  &= \frac1Z \int {\cal D}U{\cal D}\psi{\cal D}\bar\psi
  \, \hat{O}(U) \, e^{-S_g-S_q}
  \nonumber \\
  &= \frac1Z \int {\cal D}U  \, \hat{O}(U) [{\rm det} M(\kappa)]^{N_{\rm f}} e^{-S_g}
  \nonumber \\
  &= \frac1Z \int {\cal D}U \, \hat{O}(U) \, e^{-S_g+N_{\rm f} \ln{\rm det}M(\kappa)},
  \label{eq:<O>}
\end{align}
with the partition function $Z=\int {\cal D}U \, e^{-S_g+N_{\rm f} \ln{\rm det}M(\kappa)}$.

In the heavy quark limit $\kappa=0$ ($m_q = \infty$), the deconfinement phase transition
at nonzero temperature is characterized by the spontaneous symmetry
breaking of the global $Z(3)$ center symmetry of the $SU(3)$ gauge symmetry.
The most conventional choice for the order parameter of this phase transition
is the Polyakov loop
\begin{align}
\hat\Omega = \frac1{N_{\rm c}N_s^3}
\displaystyle \sum_{\vec{x}} {\rm tr_C} \left[ 
U_{\vec{x},4} U_{\vec{x}+\hat{4},4} U_{\vec{x}+2 \cdot \hat{4},4} 
\cdots U_{\vec{x}+(N_t -1) \cdot \hat{4},4} \right] , 
\label{eq:Pol}
\end{align}
where the summation $\sum_{\vec{x}}$ is over the spatial lattice sites on one time slice.
In the heavy quark limit, $\langle \hat\Omega \rangle=0$
below the critical temperature $T_{\rm c}$, while 
$\langle \hat\Omega \rangle$ takes a nonzero value at $T>T_{\rm c}$.
For finite $m_q$, the $Z(3)$ symmetry is explicitly broken by the quark term, 
and thus $\langle \hat\Omega \rangle$ becomes non vanishing for all $T$.
Even in this case, when $m_q$ is sufficiently large, $\langle \hat\Omega \rangle$ 
jumps discontinuously at the first-order transition and thus can be used to detect the first-order transition line and its CP~\cite{Saito:2011fs,Saito:2013vja,Ejiri:2019csa}.

In Sec.~\ref{sec:p(O)}, we study the scaling property of the distribution function
of the real part of the Polyakov loop
\begin{align}
  \hat\Omega_{\rm R} = {\rm Re}\,\hat\Omega ,
\end{align}
defined by
\begin{align}
  p(\Omega_{\rm R}) = \langle \delta(\Omega_{\rm R} - \hat\Omega_{\rm R}) \rangle.
  \label{eq:distr}
\end{align}
In Sec.~\ref{sec:quality}, we also calculate the double distribution function of $\hat{P}$ and $\hat\Omega_{\rm R}$ defined by 
\begin{align}
  p(P,\Omega_{\rm R})
  = \left\langle \delta(P-\hat{P}) \delta(\Omega_{\rm R}-\hat{\Omega}_{\rm R}) \right\rangle. 
  \label{eq:p(P,O)}
\end{align}
In numerical calculation of these distribution functions, because the statistics of the data is finite, we have to replace the delta functions by smeared ones with finite width.
A conventional choice for this is the normalized Gauss function, 
$\delta(x) \simeq \exp [ -(x/\Delta)^2 ] / (\Delta \sqrt{\pi})$~\cite{Saito:2011fs,Saito:2013vja}.
The width $\Delta$ should be large enough to have statistically meaningful number of data at each point within the width, 
and simultaneously small enough to resolve the functional shape of the distribution function.
Examining the resolution and the statistical error of distribution functions, we adopt $\Delta_{\Omega_{\rm R}}=0.002$ and $\Delta_{P}=0.0001$ in this study.
In Appendix~\ref{sec:smear}, we confirm that the resulting distribution functions as well as other results of observables discussed in this study are stable under variations of the widths around these values.

\subsection{Hopping parameter expansion}
\label{sec:HPE}

\begin{figure}[tb]
  \centering
  \includegraphics[width=0.4\textwidth]{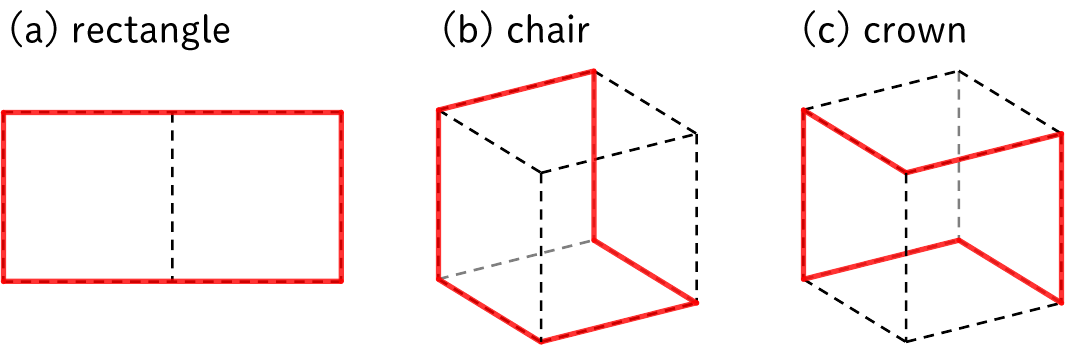}
  \caption{Six-step Wilson loops; rectangle (a), chair-type (b) and crown-type (c).}
  \label{fig:6step}
\end{figure}

\begin{figure}[tb]
  \centering
  \includegraphics[width=0.4\textwidth]{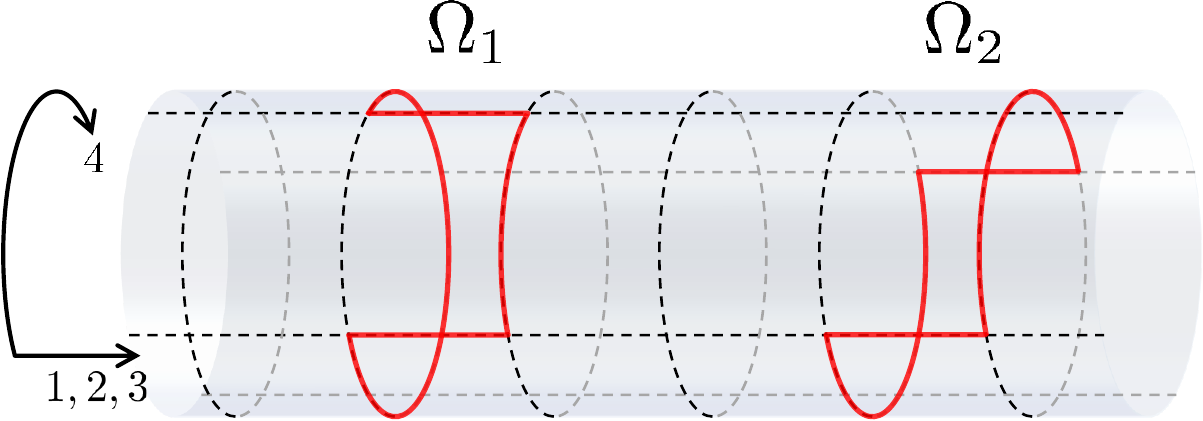}
  \caption{Bent Polyakov loops,  $\Omega_1$ (left) and $\Omega_2$ (middle)
    for $N_t=4$.}
  \label{fig:bent}
\end{figure}

To calculate Eq.~(\ref{eq:<O>}) around the heavy-quark limit,
in the present study we adopt the hopping parameter expansion (HPE)
for $\ln{\rm det}M(\kappa)$:
\begin{align}
  \ln \left[ \frac{\det M(\kappa)}{\det M(0)} \right]
  =-\sum_{n=1}^{\infty} \frac1n {\rm Tr} 
  \left[ B^n \right] \kappa^n,
  \label{eq:HPE}
\end{align}
where the matrix $B_{xy}$ is defined by Eq.~(\ref{eq:B}) and 
${\rm Tr}$ is the trace over all indices. 
In Eq.~(\ref{eq:HPE}), the contribution at $\kappa=0$ is subtracted for convenience.
Since $B_{xy}$ takes nonzero values only for adjacent
lattice sites $x$ and $y$, $B^n$ is graphically represented
by trajectories of $n$ links~\cite{Rothe:1992nt}.
Because of the trace in Eq.~(\ref{eq:HPE}),
non-vanishing contributions are given by closed
trajectories.
The lowest-order contributions of Eq.~(\ref{eq:HPE}) start from $n=4$,
and all contributions for odd $n$ vanishes when $N_t$ is even.
By writing 
\begin{align}
  -\ln \left[ \frac{\det M(\kappa)}{\det M(0)} \right]
  =& S_{\rm LO} + S_{\rm NLO} + {\cal O}(\kappa^8),
  \label{eq:HPE=S}
\end{align}
one finds for $N_t=4$ that the LO contribution is given by
the plaquette and the Polyakov loop as
\begin{align}
  S_{\rm LO} = -2N_{\rm c} \big( 48 N_{\rm site} \hat{P} + 32 N_s^3 \hat\Omega_{\rm R} \big) \kappa^4.
  \label{eq:LO}
\end{align}
The NLO term $S_{\rm NLO}$ consists of
the six-step Wilson loops and bent Polyakov loops as
\begin{align}
  S_{\rm NLO}
  =& -2N_{\rm c} \big(
  384\,\hat{W}_{\rm rec} + 768\,\hat{W}_{\rm chair} + 256\,\hat{W}_{\rm crown}
  \nonumber \\
  &+ 192 \,{\rm Re}\,\hat\Omega_1 + 96 \,{\rm Re}\,\hat\Omega_2 \big) 
  N_{\rm site} \kappa^6.
  \label{eq:NLO}
\end{align}
Here,
$\hat{W}_{\mathrm{rec}}$, $\hat{W}_{\mathrm{chair}}$, and $\hat{W}_{\mathrm{crown}}$
represent the six-step Wilson loops of the rectangular,
chair, and crown types, respectively, as illustrated in Fig.~\ref{fig:6step}.
$\hat\Omega_n$ are the bent Polyakov loops illustrated in Fig.~\ref{fig:bent},
which run one step in a spatial direction, $n$ steps in the temporal direction
and return to the original line.
All the Wilson loops and Polyakov-loop-type loops are normalized such that 
$\hat{W}_{\mathrm{rec}} = \hat{W}_{\mathrm{chair}} = \hat{W}_{\mathrm{crown}}=1$ and 
$\hat\Omega_n = 1$ in the weak coupling limit, $U_{x,\mu}=1$.
Explicit definitions of these operators as well as the derivation of
Eqs.~(\ref{eq:LO}) and (\ref{eq:NLO}) are given
in Appendix~\ref{sec:HPEapp}.

\subsection{Numerical implementation with HPE}
\label{sec:HPE-num}

In this study, we generate the gauge configurations
with respect to the action at the LO in the HPE, \textit{i.e.}
\begin{align}
  S_{g+\rm LO} 
  &= S_g+ N_{\rm f} S_{\rm LO}
  \nonumber \\
  &= -6 N_{\rm site} \big( \beta + 16 N_{\rm c} N_{\rm f} \kappa^4 \big)\hat{P}
  - 64 N_{\rm c} N_{\rm f} N_s^3 \kappa^4 \hat\Omega_{\rm R}
  \nonumber \\
  &= -6 N_{\rm site} \beta^* \hat{P}
  - \lambda N_s^3 \hat\Omega_{\rm R},
  \label{eq:g+LO}
\end{align}
with
\begin{align}
  \beta^* = \beta + 16 N_{\rm c} N_{\rm f} \kappa^4 ,
  \qquad
  \lambda = 64 N_{\rm c} N_{\rm f} \kappa^4 .
\end{align}
We then perform the measurements at the NLO by 
incorporating the effect of $S_{\rm NLO}$ by the multipoint reweighting method~\cite{Ferrenberg:1989ui,Saito:2013vja,Iwami_2015}.
In this subsection we discuss the numerical implementation of these analyses.

\begin{figure}
  \centering
  \begin{tabular}{ll}
    (a) $X_\mu(x)$ & (b) $Y_0(x)$ \vspace{2mm} \\
    \hspace{2mm}
    \includegraphics[width=0.13\textwidth]{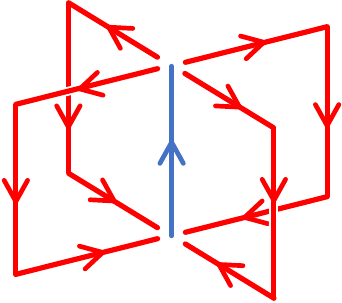}
    \hspace*{4mm} &
    \includegraphics[width=0.18\textwidth]{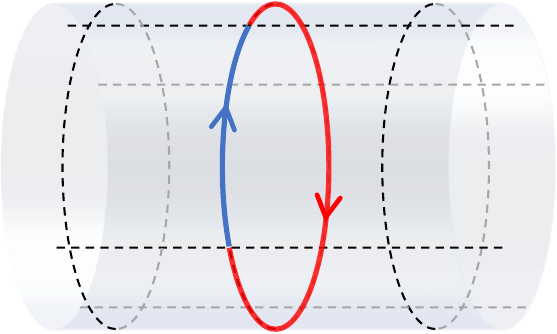}
  \end{tabular}
  \caption{
    Staple $X_\mu(x)$ and the operator $Y_0(x)$ in Eq.~(\ref{eq:Y0})
    corresponding to the link variable $U_\mu(x)$ shown by the blue
    line.
  }
\label{fig:staple}
\end{figure}

In the Monte Carlo simulations of pure gauge theory,
thanks to the locality of the action $S_g$, it is possible to 
adopt the pseudo-heat-bath (PHB) and over-relaxation (OR) algorithms 
for updating gauge configurations.
Focusing on a link variable $U_\mu(x)$,
the dependence of $S_g$ on $U_\mu(x)$ is given by 
\begin{align}
  \Delta S_g ( U_\mu(x) ) = -\frac\beta{N_{\rm c}} {\rm Re} \ {\rm tr_C} [ U_\mu(x) X_\mu(x) ],
  \label{eq:UX}
\end{align}
where the staple
\begin{align}
  X_\mu(x) = \sum_{\nu\ne\mu} \sum_{s=\pm1}
    U_{x+\hat\mu,s\nu} U_{x+s\hat\nu,\mu}^\dagger U_{x,s\nu}^\dagger,
    \label{eq:staple}
\end{align}
with $U_{x,-\mu}=U^\dagger_{x-\hat{\mu},\mu}$ for $\mu>0$, 
is graphically shown in Fig.~\ref{fig:staple} (a).
In the PHB and OR algorithms, the link variable $U_\mu(x)$ is updated
according to the probability determined by Eq.~(\ref{eq:UX}).
The fact that Eq.~(\ref{eq:UX}) is represented only by 
local variables near $U_\mu(x)$ 
makes this procedure efficient
especially on the memory-distributed parallel computing.

When the LO term, Eq.~(\ref{eq:LO}), is included into the action,
the contribution of a temporal link $U_0(x)$ to $S_{g+ \rm LO}$ is
modified as
\begin{align}
  &\Delta S_{g+\rm LO} ( U_\mu(x) )
  \nonumber \\
  &=  -\frac{\beta^*}{N_{\rm c}} {\rm Re} \ {\rm tr_C} \Big[ U_0(x) \Big( X_0(x)
  + \frac\lambda{\beta^*} Y_0(x) \Big) \Big]
  \label{eq:U(X+Y)}
\end{align}
with
\begin{align}
  Y_0(x) = U_0(x+\hat0) U_0(x+2\cdot\hat0) U_0(x+3\cdot\hat0),
  \label{eq:Y0}
\end{align}
which is schematically shown in Fig.~\ref{fig:staple}~(b).
For $N_t>4$, $Y_0(x)$ is given by the product of
$N_t-1$ link variables along the temporal direction.
The contribution of a spatial link to $S_{g+\rm LO}$
is unchanged from Eq.~(\ref{eq:UX}).

These results on $\Delta S_{g+\rm LO} ( U_\mu(x) )$ suggest that 
the Monte Carlo updates of $U_\mu(x)$ can be performed by the PHB and OR
efficiently even for $S_{g+\rm LO}$,
provided that the temporal direction is not separated
into different parallel nodes and $Y_0(x)$ can be calculated
efficiently.
Satisfying this condition is not difficult to attain for large-volume
simulations.
By taking this advantage, in this study
we perform update of gauge fields at the LO
by combining PHB and OR\footnote{
  After finishing our numerical analyses,
  we knew that a similar idea is suggested
  in Ref.~\cite{Hasenfratz:1983ce}.
  We thank F.~Karsch for notifying this literature.
}.
Compared with the pure-gauge simulation,
the increase of the numerical cost to deal with $S_{g+\rm LO}$
in this method is small since the additional multiplications
of SU(3) matrices required for an update are only $N_t-2$ times
for the temporal links and the cost to update the spatial links is unchanged.

In the measurement of observables, we include all the contribution of
$S_{\rm NLO}$
using the multipoint reweighting method\footnote{
  In Ref.~\cite{Ejiri:2019csa}, the effect of $S_{\rm NLO}$ is included
  in part effectively by the effective NLO method.
  In this study, we deal with it exactly.}.
The expectation value of a gauge observable $\hat{O}(U)$
at the NLO at the parameter set $(\beta,\kappa)$ is given from
the LO simulations at $(\tilde\beta,\tilde\kappa)$ as
\begin{align}
  \lefteqn{\langle \hat{O}(U) \rangle^{\rm NLO}_{\beta,\kappa}  }
  \nonumber\\
  &= \frac{
    \int {\cal D}U \, \hat{O}(U) \, e^{-S_{g+\rm LO}(\beta,\kappa)-S_{\rm NLO}(\beta,\kappa)}
  }{
    \int {\cal D}U \, e^{-S_{g+\rm LO}(\beta,\kappa)-S_{\rm NLO}(\beta,\kappa)} }
  \nonumber \\
  &= \frac{
    \int {\cal D}U \, \hat{O}(U) \,
    e^{-\delta S_{g+\rm LO}-S_{\rm NLO}(\beta,\kappa)}
    e^{-S_{g+\rm LO}(\tilde\beta,\tilde\kappa)}
  }{
    \int {\cal D}U \,
    e^{-\delta S_{g+\rm LO}-S_{\rm NLO}(\beta,\kappa)}
    e^{-S_{g+\rm LO}(\tilde\beta,\tilde\kappa)}
  }
  \nonumber \\
  &= \frac{
    \langle \hat{O}(U) \, e^{-\delta S_{g+\rm LO}-S_{\rm NLO}(\beta,\kappa)} \rangle^{\rm LO}_{\tilde\beta,\tilde\kappa}
  }{
    \langle e^{-\delta S_{g+\rm LO}-S_{\rm NLO}(\beta,\kappa)} \rangle^{\rm LO}_{\tilde\beta,\tilde\kappa}
  },
  \label{eq:reweight}
\end{align}  
with 
\begin{align}
  \delta S_{g+\rm LO}
  =& S_{g+\rm LO}(\beta,\kappa) - S_{g+\rm LO}(\tilde\beta,\tilde\kappa)
  \nonumber \\
  =& -6 N_{\rm site} \big( \beta-\tilde\beta + 16 N_{\rm c} N_{\rm f} (\kappa^4-\tilde\kappa^4 ) \big)\hat{P}
  \nonumber \\
  &+ 64 N_{\rm c} N_{\rm f} N_s^3 ( \kappa^4 - \tilde\kappa^4 ) \hat\Omega_{\rm R},
  \label{eq:DeltaS}
\end{align}
and $\langle \cdot \rangle^{\rm LO}_{\beta,\kappa}$ is the expectation value
taken with the action $S_{g+\rm LO}(\beta,\kappa)$.
In short, we generate gauge configurations for the LO action 
$S_{g+\rm LO}$ at several values of $(\tilde\beta,\tilde\kappa)$,
and evaluate the expectation values to the NLO at various $(\beta,\kappa)$
by the multipoint reweighting method.

\subsection{Simulation parameters}
\label{sec:simulation}

\begin{table}[bthp]
  \centering
  \caption{Simulation parameters: lattice size $N_s^3\times N_t$, $\beta^*$, and $\lambda$. 
  The value of $\kappa$ corresponding to each $\lambda$ is also listed for the case $N_{\rm f}=2$. 
  The last column is for the integrated autocorrelation time estimated using $\Omega_{\rm R}$.
  }
  \label{tab:params}
  \begin{tabular}{lllll}
    \hline
    lattice size & $\beta^*$ & $\lambda$ & $\kappa^{N_{\rm f}=2}$ & $\tau_{\rm int}$ \\ \hline
    $48^3 \times 4$ 
    & 5.6869  & 0.004 & 0.0568 & 642(150) \\
    & 5.6861  & 0.005 & 0.0601 & 837(75) \\
    & 5.6849  & 0.006 & 0.0629 & 537(49) \\
    \hline
    $40^3 \times 4$
    & 5.6885  & 0.003 & 0.0529 & 1448(160) \\
    & 5.6869  & 0.004 & 0.0568 & 685(113) \\
    & 5.6861  & 0.005 & 0.0601 & 630(60) \\
    & 5.6849  & 0.006 & 0.0629 & 416(33) \\
    & 5.6837  & 0.007 & 0.0653 & 310(24) \\
    \hline
    $36^3 \times 4$
    & 5.6885  & 0.003 & 0.0529 & 936(113) \\
    & 5.6869  & 0.004 & 0.0568 & 459(49) \\
    & 5.6861  & 0.005 & 0.0601 & 511(46) \\
    & 5.6849  & 0.006 & 0.0629 & 364(23) \\
    & 5.6837  & 0.007 & 0.0653 & 278(16) \\
    \hline
    $32^3 \times 4$
    & 5.6885  & 0.003 & 0.0529 & 646(62) \\
    & 5.6865  & 0.004 & 0.0568 & 307(34) \\
    & 5.6861  & 0.005 & 0.0601 & 401(32) \\
    & 5.6845  & 0.006 & 0.0629 & 270(18) \\
    & 5.6837  & 0.007 & 0.0653 & 225(15) \\
    \hline
    $24^3 \times 4$
    & 5.6870  & 0.0038 & 0.0561 & 464(32) \\
    & 5.6820  & 0.0077 & 0.0669 & 250(18) \\
    & 5.6780  & 0.0115 & 0.0740 & 160(10) \\
    \hline
  \end{tabular}
\end{table}

In this study, we perform Monte Carlo simulations 
with fixed temporal lattice size $N_t=4$, while the 
spatial extent $N_s$ is changed from $24$ to $48$.
This allows us to perform simulations with large aspect ratio
$N_s/N_t=LT$ up to $12$, where $L$ is the lattice size along the spatial
direction in physical units.
For each $N_s$, the gauge configurations are generated
for $3$ to $5$ sets of $(\beta^*,\lambda)$
shown in Table.~\ref{tab:params}, which are chosen 
so that $\beta^*$ is close to the transition line at the LO.
All numerical results shown in Secs.~\ref{sec:Binder} and~\ref{sec:p(O)}
are generated by the multipoint reweighting method from these configurations.

The gauge configurations are updated with the LO action $S_{g+\rm LO}$
using the PHB and OR algorithms as discussed in Sec.~\ref{sec:HPE-num}.
Gauge configurations are updated by five OR steps after each PHB step. 
We measure observables every two sets of the PHB+OR updates, \textit{i.e.} totally ten OR steps and two PHB steps.
For all parameters we have performed $6\times10^5$ measurements in this way.
In the following, we set $N_{\rm f}=2$ to show the numerical results
unless otherwise stated.

\begin{figure}
  \centering
  \includegraphics[width=0.48\textwidth]{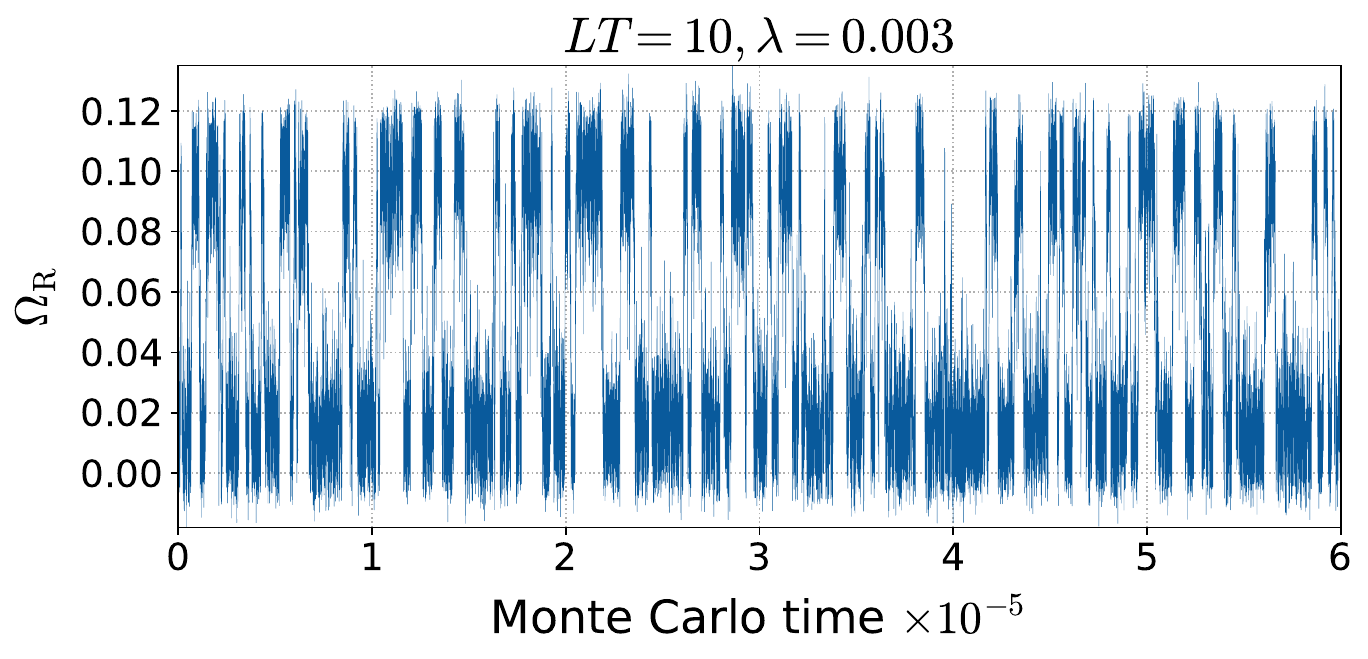}
  \includegraphics[width=0.48\textwidth]{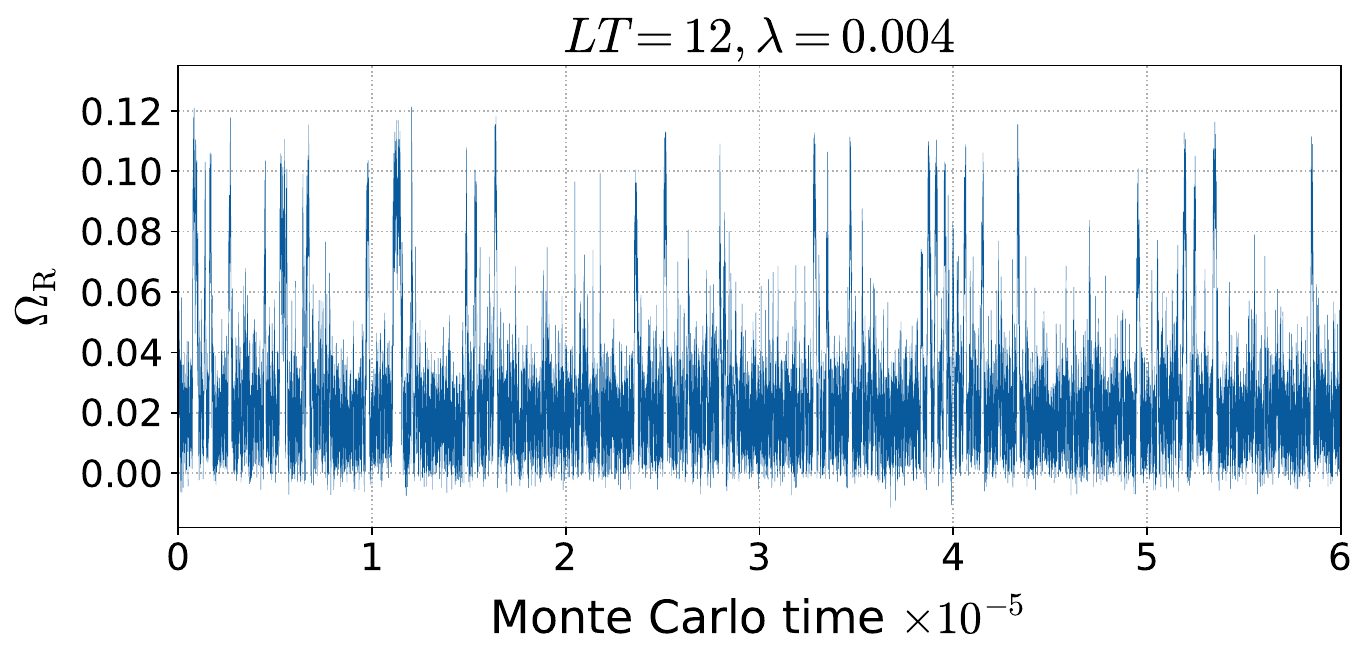}
  \caption{
    Monte Carlo time history of $\hat{\Omega}_{\rm R}$ 
    at the smallest $\lambda$ for $LT=12$ and $LT=10$.
    Horizontal axis represents the Monte Carlo time in the unit of
    measurements which are made every two sets of PHB+OR updates.
  }
  \label{fig:history}
\end{figure}

\begin{figure}
  \centering
  \includegraphics[width=0.48\textwidth]{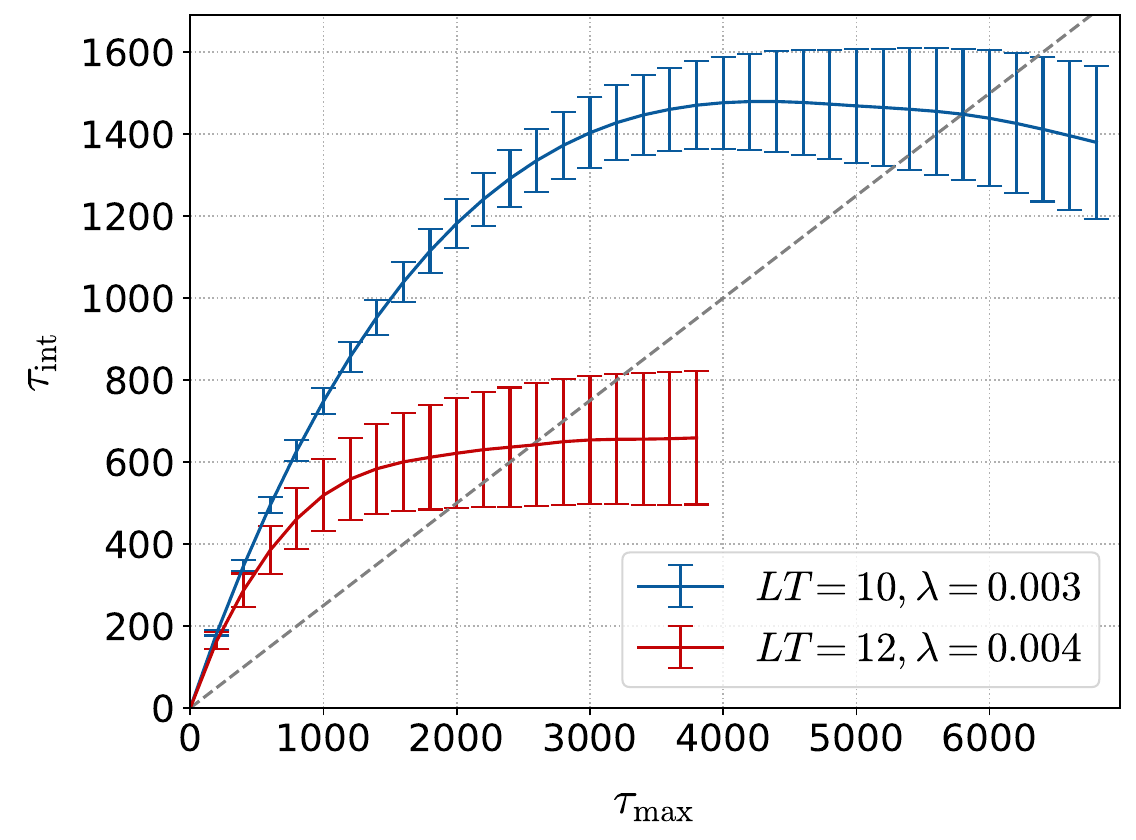}
  \caption{
    Integrated autocorrelation time $\tau_{\rm int}$
    of $\hat{\Omega}_{\rm R}$ as a function of $\tau_{\rm max}$
    on the lattices shown in Fig.~\ref{fig:history}.
    The dashed line shows $\tau_{\rm max}= 4\tau_{\rm int}$.
    }
  \label{fig:tau_int}
\end{figure}

In Monte Carlo simulations near a first-order transition,
because the transition between the coexisting two phases becomes rare when the spatial volume of the system is large, observables averaged over the two phases tend to have quite long autocorrelations.
In Fig.~\ref{fig:history} we show the Monte Carlo time history
of $\hat{\Omega}_{\rm R}$ for $N_s/N_t= LT=10$ and $12$ at the
smallest $\lambda$, at which the autocorrelation is the longest.
The horizontal axis represents the Monte Carlo time in the unit of measurements.
To estimate the autocorrelation time, 
in Fig.~\ref{fig:tau_int}, we plot 
the integrated autocorrelation time of $\Omega_{\rm R}$ defined as
\begin{align}
  \tau_{\rm int} = \frac12 + \sum_{\tau=1}^{\tau_{\rm max}}
  \frac{ \langle \Omega_{\rm R}(\tau) \Omega_{\rm R}(0) \rangle }
       { \langle \Omega_{\rm R}(0) \Omega_{\rm R}(0) \rangle },
       \label{eq:tau_int}
\end{align}
as a function of $\tau_{\rm max}$ for the two Monte Carlo trajectories
shown in Fig.~\ref{fig:history},
where $\tau$ represents the Monte Carlo time.
We estimate the autocorrelation time from the value of $\tau_{\rm int}$
at $\tau_{\rm max}\simeq 4\tau_{\rm int}$, i.e. the crossing point
between $\tau_{\rm int}$ and the dashed line in Fig.~\ref{fig:tau_int},
at which the $\tau_{\rm max}$ dependence is well saturated.
The values of $\tau_{\rm int}$ thus determined on each lattice
are listed in Table~\ref{tab:params}.

Throughout this study, we estimate the statistical errors
of observables by the jackknife method unless otherwise stated, 
adopting the binsize of $10,000$ measurements which is sufficiently larger than the estimated autocorrelation lengths.
We checked that the statistical errors thus estimated are roughly stable within a variation of the binsize from $5,000$ to $30,000$ measurements.

\begin{figure}
  \centering
  \includegraphics[width=0.48\textwidth]{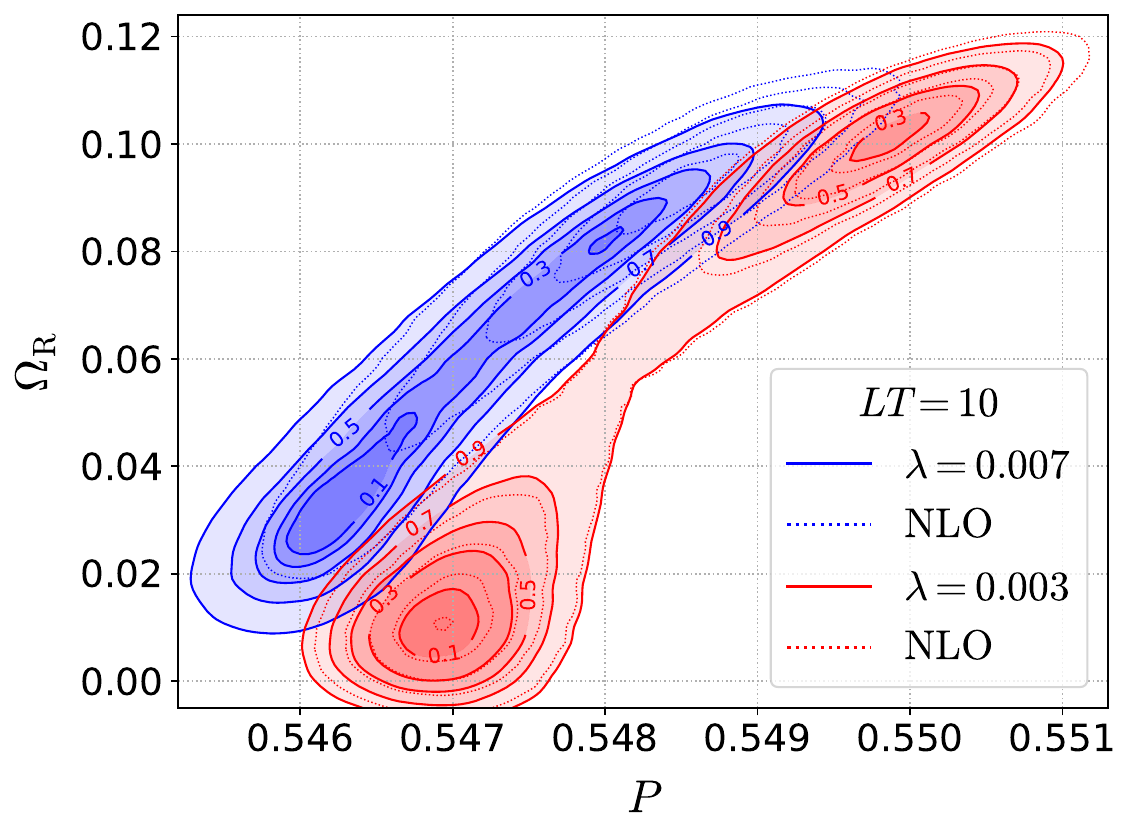}
  \caption{
    Probability distribution $p(P,\Omega_{\rm R})$ of
    the gauge configurations on the $40^3\times4$ lattice
    for the simulation points at $\lambda=0.003$ and $0.007$.
    Solid lines represent the contours of $p(P,\Omega_{\rm R})$
    obtained with the LO action.
    The label on each contour shows the probability inside the contour.
    Dotted lines represents the contour lines of $p(P,\Omega_{\rm R})$ with the NLO action
    at the same parameters obtained by reweighting the LO data.
  }
\label{fig:contour}
\end{figure}

\subsection{Overlapping problem}
\label{sec:quality}

Our main objective to use the LO action for configuration generation is to avoid the overlapping problem in the reweighting.
In Fig.~\ref{fig:contour} we show the contour plot for the
probability distribution function 
$p(P,\Omega_{\rm R})$ defined by Eq.~(\ref{eq:p(P,O)}),
obtained at $\lambda=0.003$ (red) and $0.007$ (blue) on $40^3\times4$ lattices
adopting $\Delta_{\Omega_{\rm R}}=0.002$ and $\Delta_P=0.0001$ for the smearing widths.
We checked that $p(P,\Omega_{\rm R})$ hardly changes under
variation of $\Delta_{\Omega_{\rm R}}$ and $\Delta_P$ around this choice.
The solid lines are the contours for the distribution measured on 
the LO configurations generated with $S_{g+\rm LO}$.
Each contour curve is drawn such that the probability
inside the contour is $0.9$, $0.7$, $\cdots$, and $0.1$.

In Fig.~\ref{fig:contour}, we also show by the dotted lines the contours of $p(P,\Omega_{\rm R})$ at the NLO, which is obtained by reweighting the LO data at the same $(\beta^*,\lambda)$.
The meaning of the contours is the same as the solid lines.
We find that the deviation of the NLO distribution from the
original one at the LO is not significant, suggesting that the effects of the NLO contribution are not large.
The large overlap of the LO and NLO distributions ensures that, for observables constructed from
$\hat{P}$ and $\hat{\Omega}_{\rm R}$, the NLO results obtained by reweighting the LO data at the same $\beta^*$ and $\lambda$ are statistically reliable.
In the analysis of the CP, the overlapping of the distributions
is even more improved after adjusting the parameters to the transition line.

On the other hand, from this figure, we find that the overlapping of the distributions at $\lambda=0.003$ and $0.007$ is quite poor --- the regions with probability larger than 0.7 are not overlapping at all with each other.
This means that, if we were to calculate observables at $\lambda=0.007$ by reweighting data obtained at $\lambda=0.003$, or vice versa, the statistical quality of the results would be quite low.
In Refs.~\cite{Saito:2011fs,Saito:2013vja,Ejiri:2019csa},
the CP in heavy quark region was investigated on lattices with $N_s/N_t=4$--$6$, by reweighting from pure gauge
configurations, \textit{i.e.} those obtained at $\lambda=0$.
Because the overlapping problem becomes quickly severe as the system volume becomes large, 
the same strategy is not applicable to the present study in which much larger system volumes up to $N_s/N_t=LT=12$ are simulated.
Figure~\ref{fig:contour} shows that, to incorporate the NLO effects by reweighting, 
the use of the LO action for configuration generation is sufficiently effective in suppressing the overlapping problem.  
The smallness of the NLO effects in Fig.~\ref{fig:contour} further suggests that the effects of dynamical quarks are dominated by the LO term for these values of $\lambda$.

\section{Binder cumulant analysis}
\label{sec:Binder}

\subsection{Transition line}
\label{sec:beta_c}

\begin{figure*}
  \centering
    \includegraphics[width=0.325\textwidth]{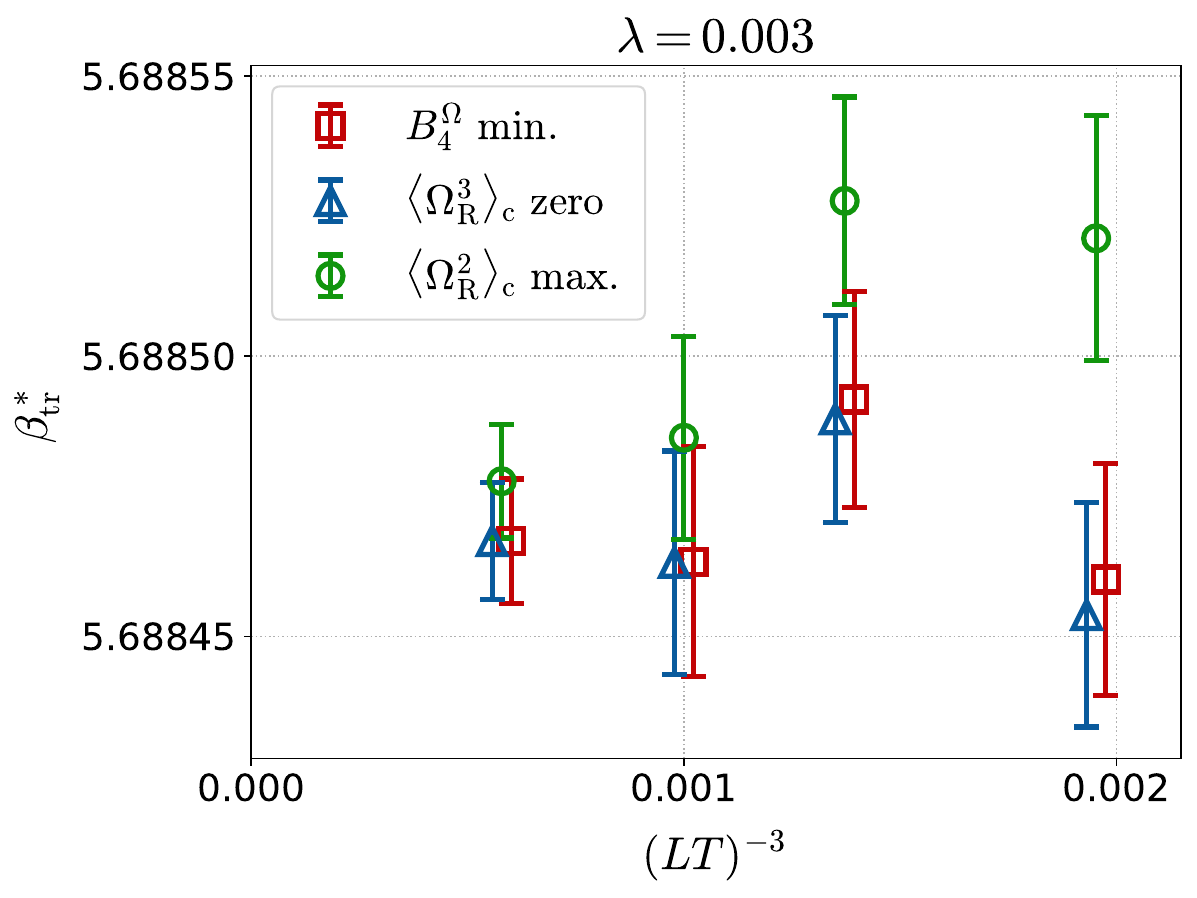}
    \includegraphics[width=0.325\textwidth]{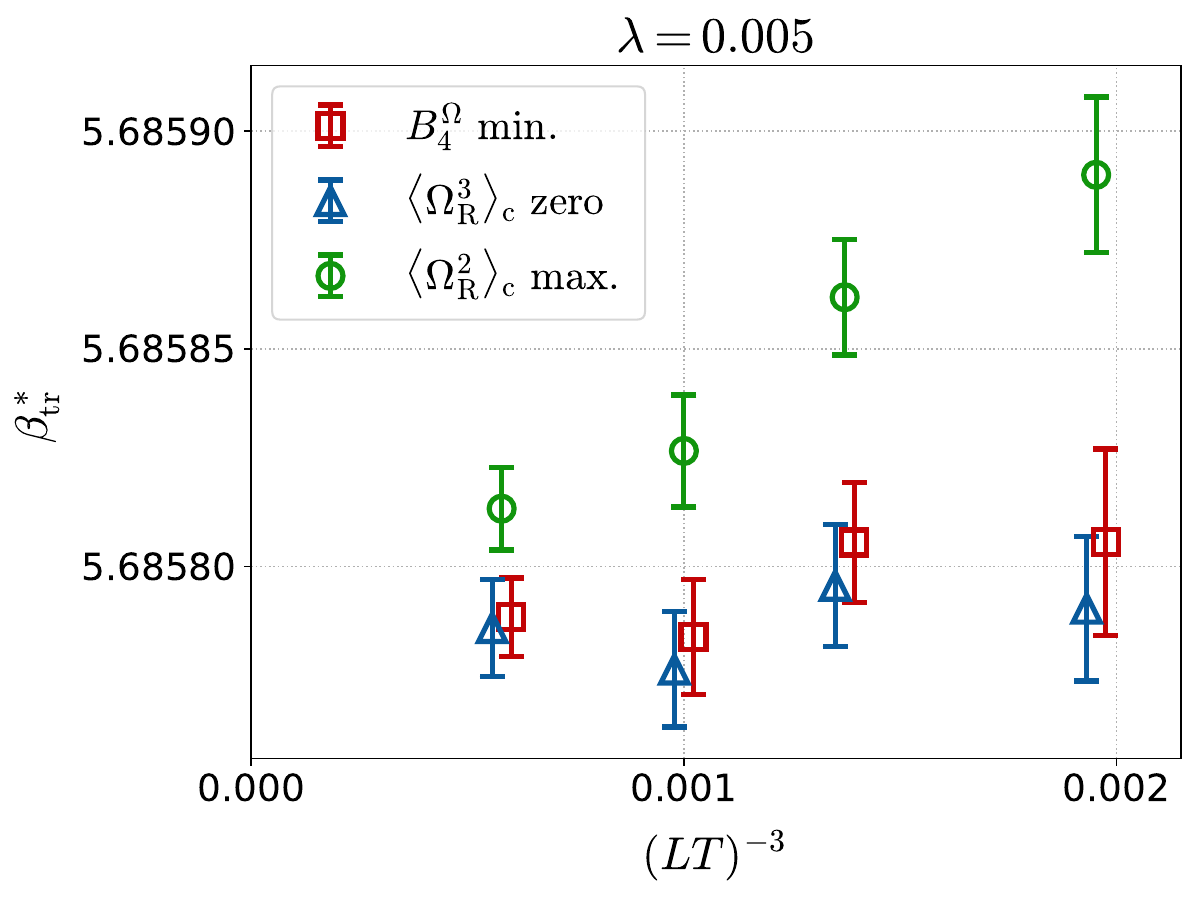}
    \includegraphics[width=0.325\textwidth]{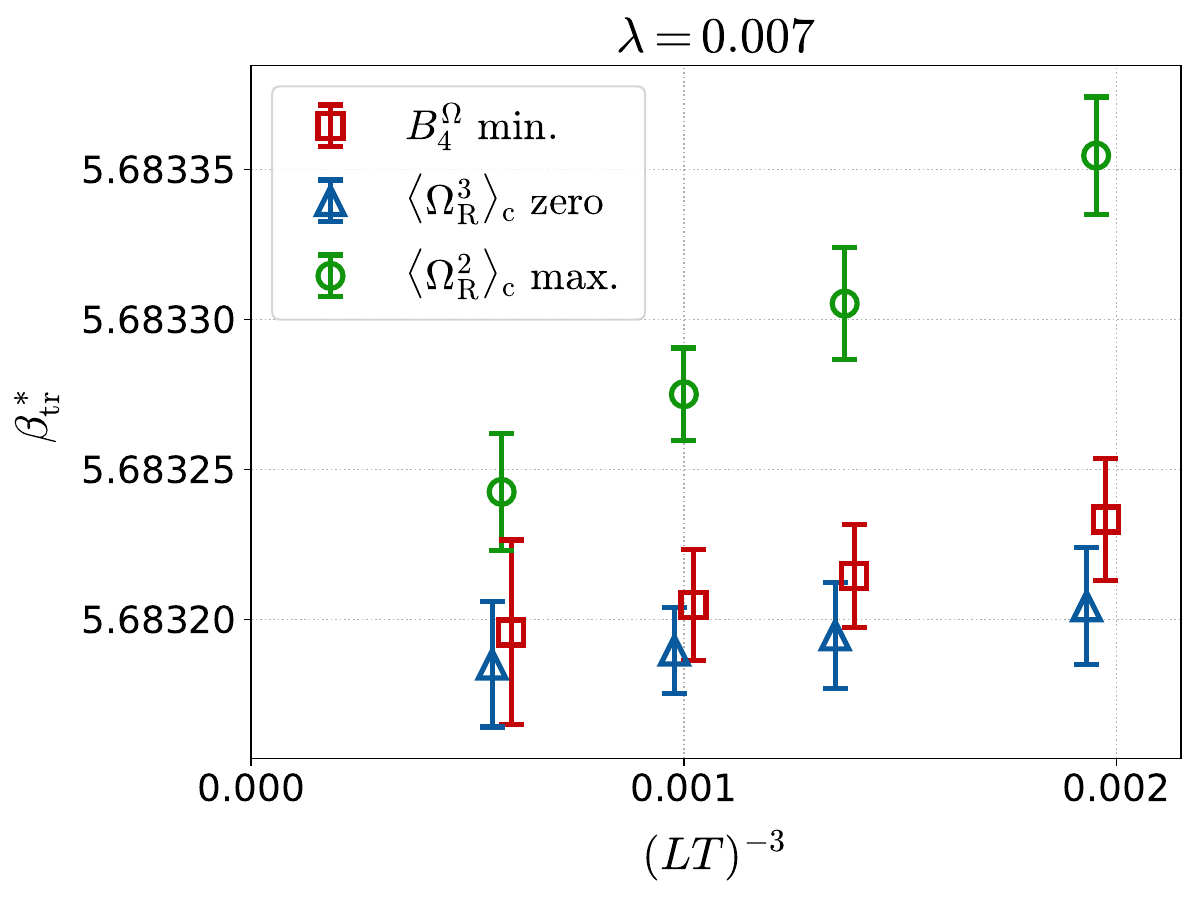}
  \caption{
    Transition line $\beta^*_{\rm tr}(\lambda)$ as a function of the aspect ratio $LT=N_s/N_t$, determined through
    three definitions:
    (1) minimum of $B_4^\Omega$, (2) zero of $\langle \Omega_{\rm R}^3 \rangle_{\rm c}$,
    and (3) maximum of $\langle \Omega_{\rm R}^2 \rangle_{\rm c}$.
    The results for $\lambda=0.003$, 0.005, 0.007 are shown from left to right.
  }
\label{fig:beta_c}
\end{figure*}

We first determine the location of the transition line that corresponds to $h=0$
in terms of the Ising parameters; see Fig.~\ref{fig:scaling}.
In the coupling parameter space $(\beta^*,\lambda)$,
we denote the transition line as $\beta^*_{\rm tr}(\lambda)$.
In this study,
we determine $\beta^*_{\rm tr}$ at each $\lambda$ adopting the following three conventional choices:
\begin{itemize}
\item
  Maximum of $\langle \Omega_{\rm R}^2 \rangle_{\rm c}$
\item
  Zero point of $\langle \Omega_{\rm R}^3 \rangle_{\rm c}$
\item
  Minimum of $B_4^\Omega=\langle \Omega_{\rm R}^4 \rangle_{\rm c}/\langle \Omega_{\rm R}^2 \rangle_{\rm c}^2+3$
\end{itemize}

\begin{figure}
  \centering
    \includegraphics[width=0.48\textwidth]{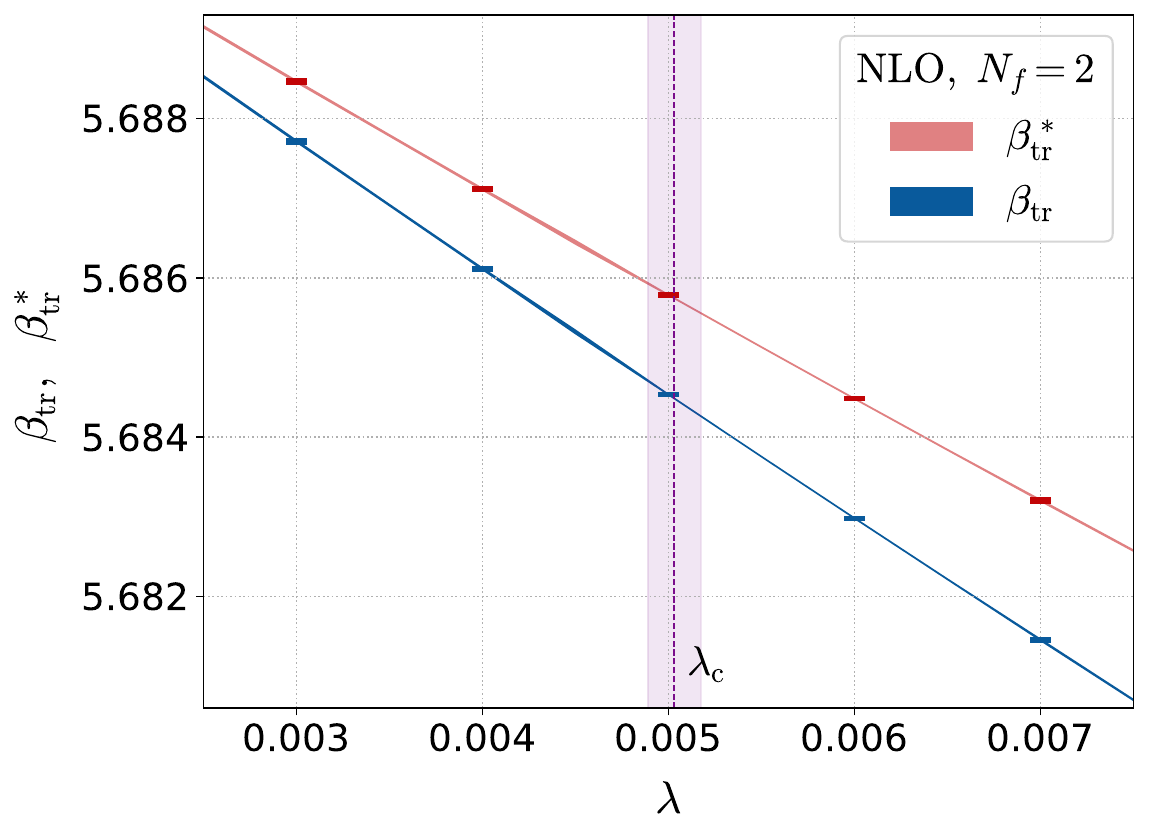}
  \caption{
    Transition line $\beta^*_{\rm tr}$ on the $(\beta,\lambda)$ and $(\beta^*,\lambda)$ plane obtained at $LT=10$.
    The points with the error bar are the results obtained on
    each measurement.
    The shaded areas show the result obtained by the
    multipoint reweighting method with the error band representing
    the statistical error.
    The dotted vertical line shows the value of $\lambda_{\rm c}$
    determined by the analysis in Sec.~\ref{sec:B4}.
  }
\label{fig:beta_lambda}
\end{figure}

In Fig.~\ref{fig:beta_c}, we show the $LT$ dependence of
$\beta^*_{\rm tr}$ determined by these definitions 
for several values of $\lambda$.
The figure shows that the maximum of $\langle \Omega_{\rm R}^2 \rangle_{\rm c}$
has a visible $LT$ dependence.
On the other hand, the zero point of 
$\langle \Omega_{\rm R}^3 \rangle_{\rm c}$ and the minimum of $B_4^\Omega$
do not have statistically-significant $LT$ dependence for $LT\ge8$.
This result shows that the zero point of $\langle \Omega_{\rm R}^3 \rangle_{\rm c}$
and minimum of $B_4^\Omega$ are sufficiently close to the $\beta^*_{\rm tr}$
in the $L\to\infty$ limit in this range of $LT$.
In the following, we employ the minimum of $B_4^\Omega$ for the
definition of the transition line $\beta^*_{\rm tr}$ for each $N_t$.
In Fig.~\ref{fig:beta_lambda}, we show the transition line on the
$(\beta^*,\lambda)$ and $(\beta,\lambda)$ planes obtained at $LT=10$.

\begin{figure*}
  \centering
    \includegraphics[width=0.325\textwidth]{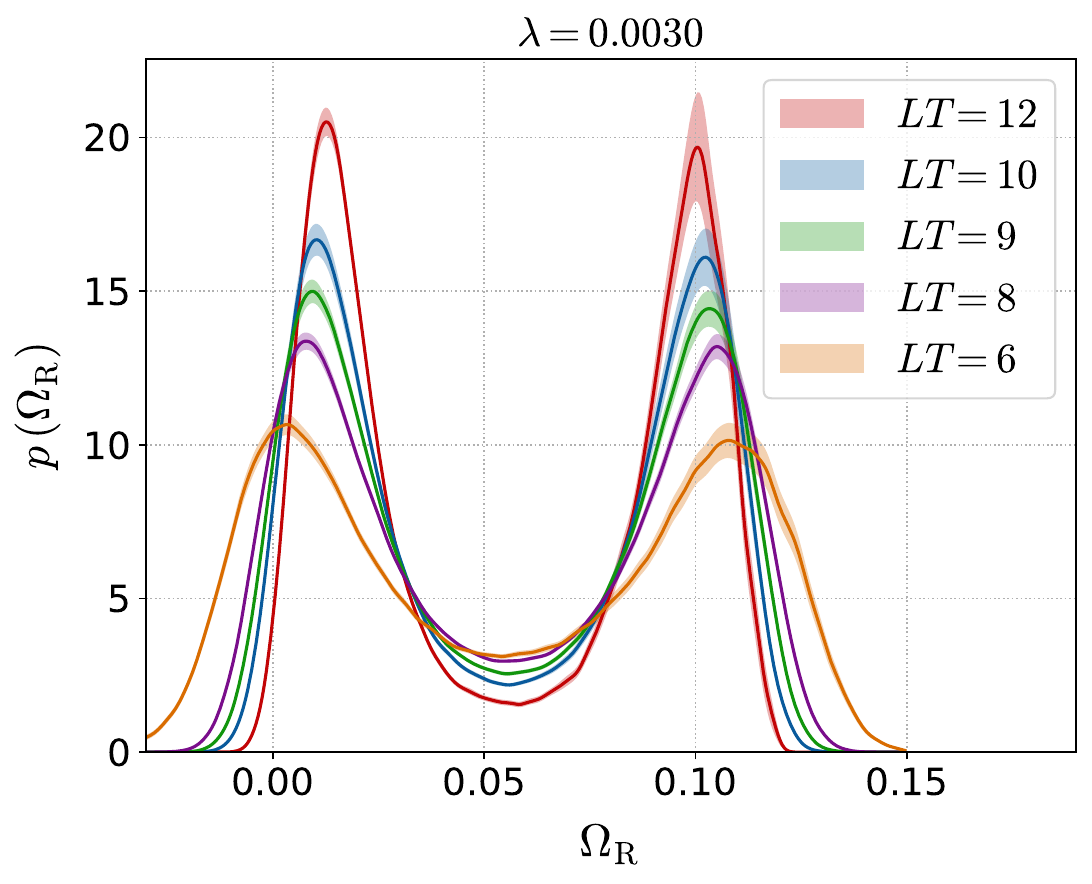}
    \includegraphics[width=0.325\textwidth]{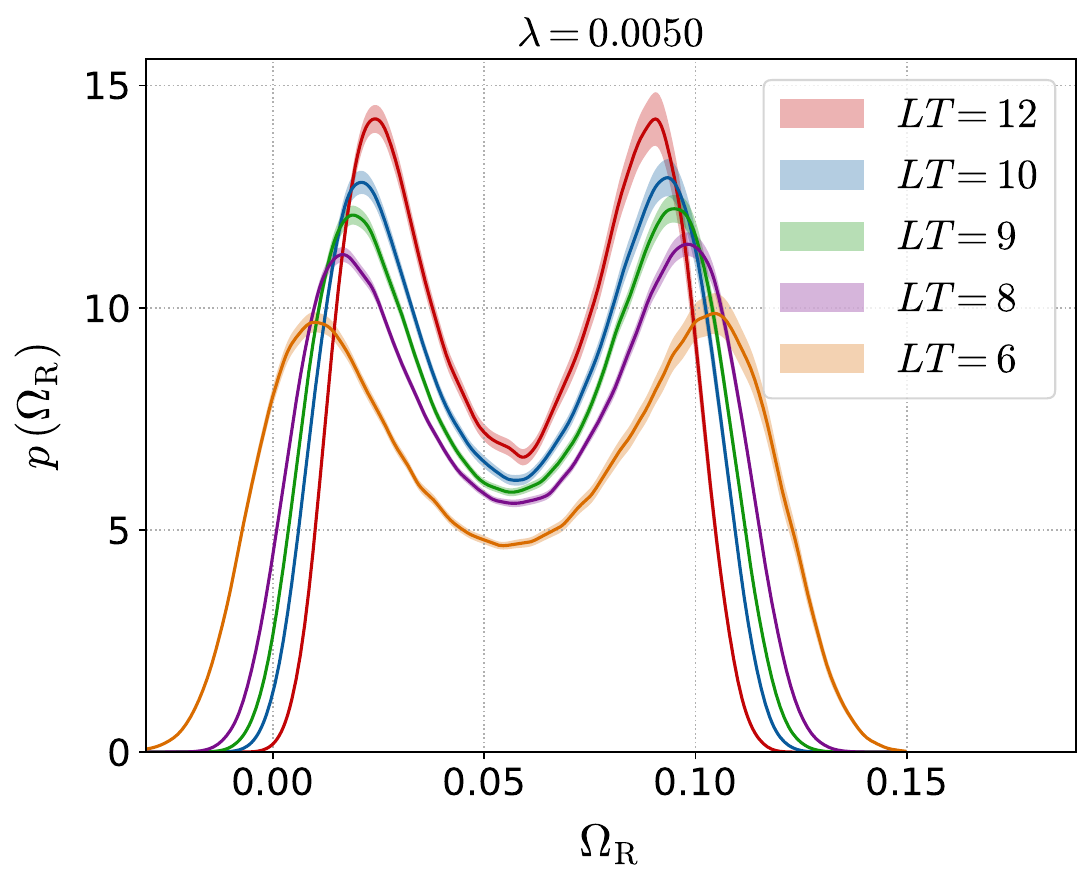}
    \includegraphics[width=0.325\textwidth]{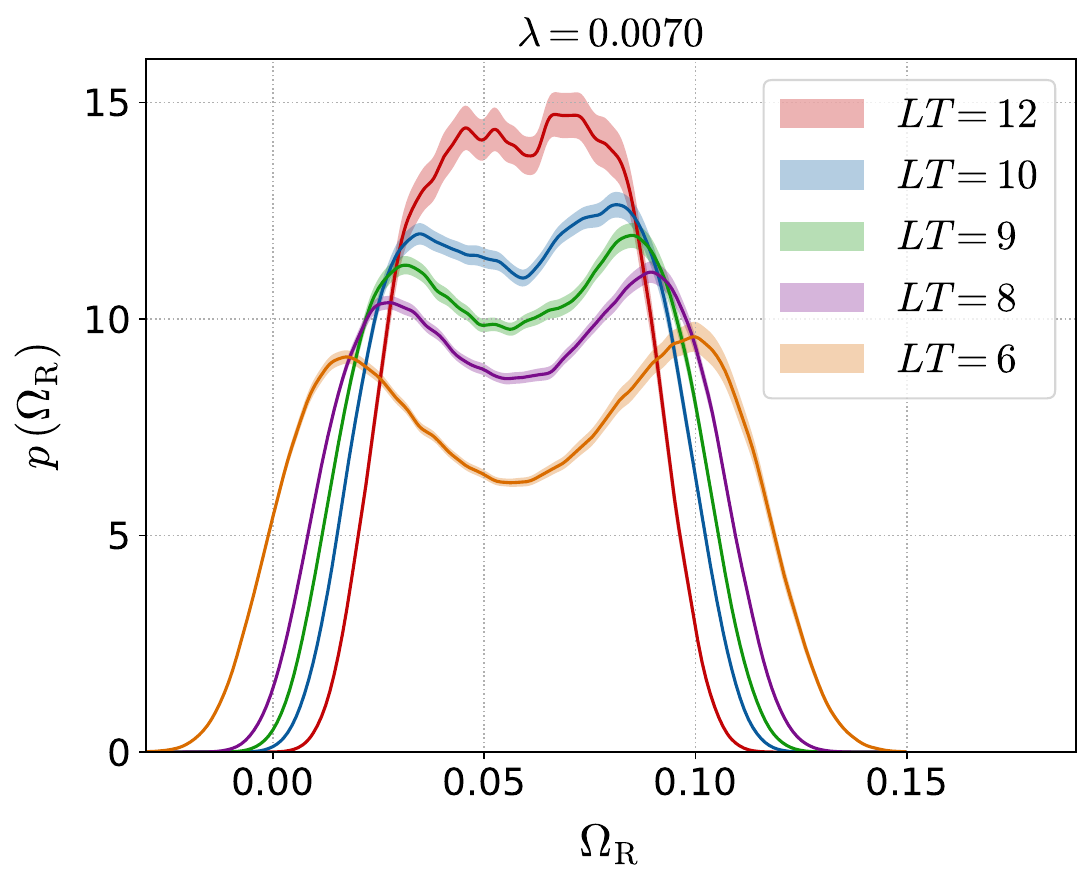}
  \caption{
    Distribution of $\Omega_{\rm R}$ at $\lambda=0.003$, $0.005$ and
    $0.007$.
  }
\label{fig:distr}
\end{figure*}

In Fig.~\ref{fig:distr}, we show 
the distribution function of $\Omega_{\rm R}$, Eq.~(\ref{eq:distr}), 
on the transition line for several values of $\lambda$, where 
the delta function in Eq.~(\ref{eq:distr}) is smeared by 
the Gauss function as before with the width
$\Delta_{\Omega_{\rm R}}=0.002$.  
As discussed in Appendix~\ref{sec:smear},
dependence of these results on $\Delta_{\Omega_{\rm R}}$ is well
suppressed around this $\Delta_{\Omega_{\rm R}}$.
The shaded bands represent the statistical errors.
At $\lambda=0.003$, we see a clear two-peak structure
in $p(\Omega_{\rm R})$ and find that the peaks become sharper
as $LT$ becomes larger.
This behavior suggests the first-order phase transition at
$\lambda=0.003$.
At $\lambda=0.007$, on the other hand, while two peaks are observed
for $LT\le9$, the two peaks cease to exist as $LT$ becomes 
large.
This suggests the crossover transition in the $L\to\infty$ limit
at this $\lambda$.

\subsection{Binder cumulant}
\label{sec:B4}

\begin{figure}
  \centering
    \includegraphics[width=0.48\textwidth]{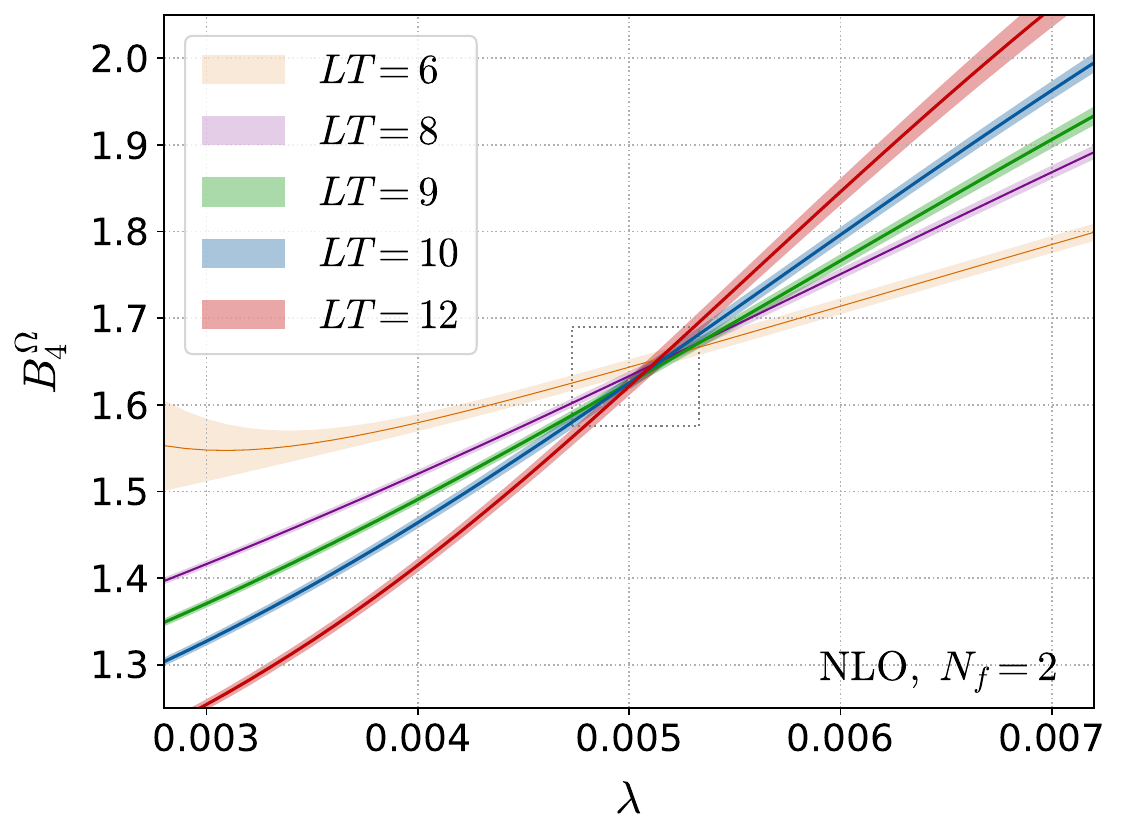}
    \includegraphics[width=0.48\textwidth]{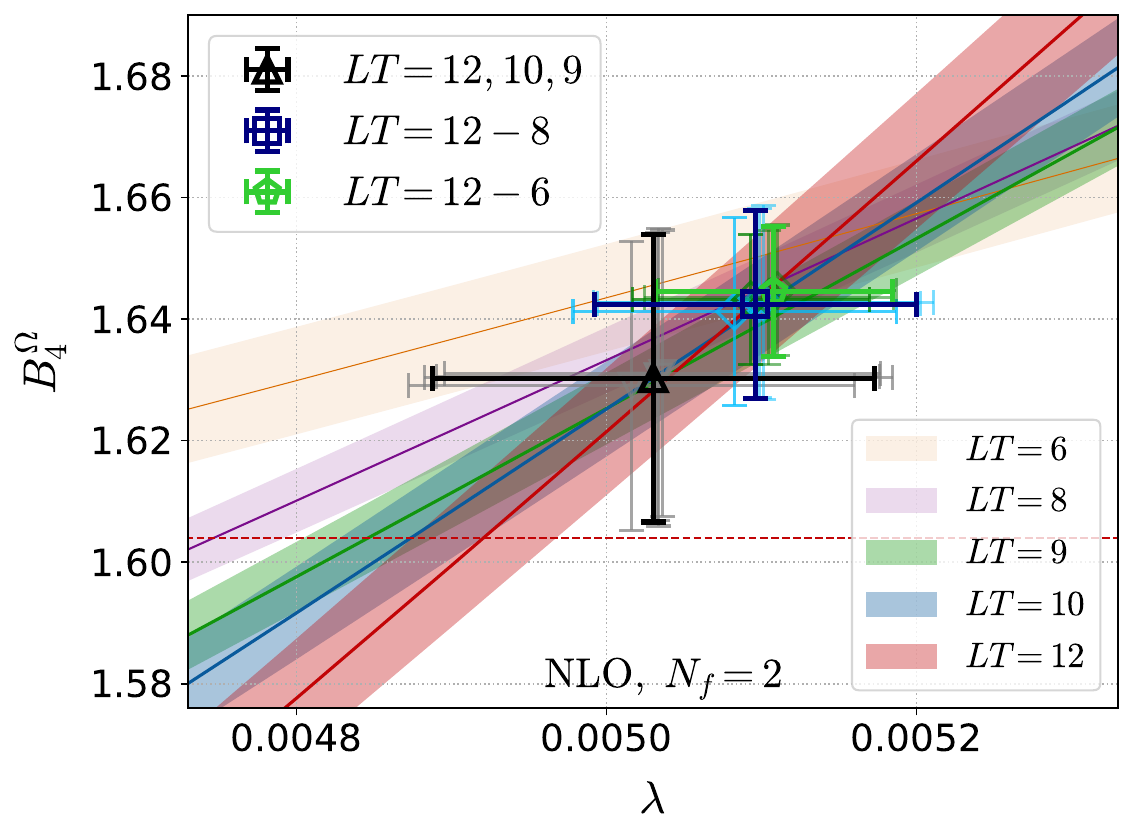}
  \caption{
    Binder cumulant $B_4^\Omega$ as a function of $\lambda$
    obtained at five $LT=N_s/N_t$. The statistical errors are shown by
    the shaded area.
    The lower panel is an enlargement of the upper panel around the crossing point, 
    where the dotted rectangle in the upper panel represents
    the region of the lower panel.
    The points in the lower panel with error bars show the
    results of the four parameter fit with Eq.~(\ref{eq:B4fit4}).
    See text for details.
  }
\label{fig:B4}
\end{figure}

Next, let us determine the position of the CP
on the $(\beta,\kappa)$ plane.
As discussed in Sec.~\ref{sec:FSS}, 
it is convenient
to employ the Binder cumulant $B_4$ of $\Omega_{\rm R}$
\begin{align}
  B_4^\Omega
  = \frac{ \langle \Omega_{\rm R}^4 \rangle_{\rm c} }
  { \langle \Omega_{\rm R}^2 \rangle_{\rm c}^2 } + 3.
  \label{eq:B4O}
\end{align}
This quantity
approaches the known values given in Eq.~(\ref{eq:B4lim}) in the $L\to\infty$ limit
depending on the order of the transition. 
Furthermore, provided that $\Omega_{\rm R}$
corresponds to $m=M/V$ of the Ising model, 
$B_4^\Omega$ should obey Eq.~(\ref{eq:B4L}) near the CP.

In the upper panel of Fig.~\ref{fig:B4}, we show 
$B_4^\Omega$ along the transition line as a function of $\lambda$
for five values of $LT$.
$\lambda$ is varied continuously by the multipoint reweighting method.
The lower panel is an enlargement of the upper panel around the crossing point.
The figure shows that $B_4^\Omega$ has a crossing at
$\lambda=\lambda_{\rm c}\simeq0.005$ and is an increasing (decreasing)
function of $LT$ for $\lambda>\lambda_{\rm c}$ ($\lambda<\lambda_{\rm c}$).
The existence of the CP at $\lambda\simeq0.005$ is suggested
from this result.

\begin{table}
  \centering
  \caption{
    Results of the four parameter fits of $B_4^\Omega$ with
    Eq.~(\ref{eq:B4fit4}).
    Three largest volumes with $LT=N_s/N_t=12,~10,~9$ are used for the fits.
    The left column shows the values of $\lambda$ used for the fit.
  }
  \label{table:B4lambda}
  \begin{tabular}{c|ccc|c}
    \hline
    fit $\lambda~(\times10^4)$ & $b_4$ & $\lambda_{\rm c}$ & $\nu$ & $\chi^2/{\rm dof}$
    \\ \hline
    49, 52 &
    1.631(24) & 0.00503(14) & 0.614(49) & 0.45 \\
    {\bf 48, 53 } &
    {\bf 1.630(24)} & {\bf 0.00503(14)} & {\bf 0.614(48)} & {\bf 0.46} \\
    45, 55 &
    1.629(24) & 0.00502(14) & 0.622(48) & 0.46 \\
    45, 50 &
    1.630(24) & 0.00503(15) & 0.634(47) & 0.48 \\
    50, 55 &
    1.631(24) & 0.00504(14) & 0.610(51) & 0.38 \\
    45, 50, 55 &
    1.620(23) & 0.00494(14) & 0.626(41) & 7.9 \\
    \hline
  \end{tabular}
\end{table}

To determine $\lambda_{\rm c}$ and the critical exponent $\nu$ 
quantitatively, 
we fit the numerical results of $B_4^\Omega$ by a fitting function motivated by Eq.~(\ref{eq:B4L}):
\begin{align}
  B_4^\Omega(\lambda,LT) = b_4 + c ( \lambda-\lambda_{\rm c} ) (LT)^{1/\nu} ,
  \label{eq:B4fit4}
\end{align}
where $b_4$, $\lambda_{\rm c}$, $\nu$, $c$ are the fit parameters. 
In this study, we can vary $\lambda$ continuously by the multipoint reweighting method.
However, because data at different $\lambda$ on the same volume are correlated, 
it is not meaningful to use too many $\lambda$ values.
Using the data at three largest volumes, $LT=12,~10,~9$, two or three $\lambda$ values
(6 or 9 data points, respectively) 
should be 
sufficient for the four parameter fit of Eq.~(\ref{eq:B4fit4}). 
We thus repeat the fit for several choices of $\lambda$ values, 
taking the covariance between data at different $\lambda$ into account in the calculation of $\chi^2$.

In Table~\ref{table:B4lambda}, we summarize the results of 
the fit using the data at three largest volumes, $LT=12,~10,~9$, 
and at $\lambda$ values listed in the left column of the table.
The statistical error in the table is estimated by the standard chi-square analysis.
The table shows that the value of $\chi^2/{\rm dof}$ are smaller than unity in the fits with two $\lambda$ values,
but $\chi^2/{\rm dof}$ is unacceptably large with three $\lambda$ values,
while all the results for the fitting parameters are well consistent within errors.
We choose the fit result for $\lambda=(0.0048,~0.0053)$ depicted by bold
characters in Table~\ref{table:B4lambda} as the central
value and include the uncertainty in the fits with two $\lambda$ values
as the systematic error.

\begin{table}
  \centering
  \caption{
  Results of the four parameter fit of $B_4^\Omega$ using the data points at three and four largest and all volumes.
  The first parentheses are for statistical errors estimated by the jackknife method, and the second parentheses are for systematic errors due to the choice of $\lambda$ values for the fit. See Sec.~\ref{sec:B4} for details.
    }
  \label{table:B4Ns}
  \begin{tabular}{c|ccc|c}
    \hline
    $LT=N_s/N_t$ & $b_4$ & $\lambda_{\rm c}$ & $\nu$ & $\chi^2/{\rm dof}$
    \\ \hline
    12, 10, 9 &
    1.630(24)(2) & 0.00503(14)(2) & 0.614(48)(3) & 0.46 \\
    12, 10, 9, 8 &
    1.643(15)(2) & 0.00510(10)(2) & 0.614(29)(3) & 0.37 \\
    12, 10, 9, 8, 6 &
    1.645(11)(2) & 0.00511(8)(2) & 0.593(18)(3) & 0.67 \\
    \hline
  \end{tabular}
\end{table}

We repeat the analysis also with other sets of system volumes. 
The results of the fits with four and five largest volumes, 
together with the fit with three largest volumes, are summarized
in Table~\ref{table:B4Ns}. 
For the fit with four and five largest volumes, $LT=12$--$8$ and $12$--$6$, 
we now have 8 and 10 data points for the four parameter fit with two $\lambda$ values.
For these fits, we choose $\lambda=(0.0048,0.0053)$ as the central value again.

The results of $b_4$ and $\lambda_{\rm c}$ obtained by these fits
are shown in the lower panel of Fig.~\ref{fig:B4}.
The thick symbols show the central value,
and thin symbols with light colors show the results obtained by
the variation of $\lambda$ in Table~\ref{table:B4lambda}.
The results with the three largest volumes are shown by black triangles, while those with four and five largest volumes are shown by blue squares and green pentagons, respectively.
In the figure, $b_4$ expected from the $Z(2)$ universality, Eq.~(\ref{eq:Z2b4}), 
is shown by the dashed horizontal line.

From Table~\ref{table:B4Ns} and Fig.~\ref{fig:B4},
we find that, when we adopt the fit with the three largest volumes, $LT\ge9$, the fit result of $b_4$ is consistent
with the $Z(2)$ value within about $1\sigma$.
On the other hand, when we include smaller volumes, $LT\ge8$ or $LT\ge6$, 
$b_4$ from the fits show statistically significant deviation from the $Z(2)$ value.
In Table~\ref{table:B4Ns}, we also summarize the results for the critical exponent $\nu$.
From the $Z(2)$ universality class, we expect $\nu=1/y_t \approx 0.630$.
We find that the result of the fit with $LT\ge9$ is consistent with the $Z(2)$ value within the error, 
while the result of the fit with $LT\ge6$ has a significant deviation from the $Z(2)$ value, 
though the values of $\chi^2/{\rm dof}$ are all smaller than unity.

We thus conclude that the FSS in the $Z(2)$ universality class
is confirmed when the system volume is large enough, $LT\ge9$
--- lattices with $LT\le8$ are not large enough
to apply an FSS analysis for $B_4^\Omega$.
The value of $\lambda_{\rm c}$ thus determined is also shown
in Fig.~\ref{fig:beta_lambda}.

In Appendix~\ref{sec:LO}, we perform the analysis of $B_4^\Omega$
at the LO of the HPE
and compare the results with those at the NLO discussed in this Section.
We find that
the LO result for $\lambda_{\rm c}$ is about 2.6\% larger than the NLO value. 
This small difference suggests that the truncation error 
of the HPE is well under control at the NLO around $\lambda_{\rm c}$.

\subsection{Mixing with energy-like observable}

So far, we have performed the analyses of $B_4^\Omega$ assuming that
$\Omega_{\rm R}$ corresponds to the magnetization $m=M/V$
of the Ising model.
Although our numerical results thus far are
in good agreement with this assumption, a possible mixing with
the energy-like observable~\cite{Karsch:2001nf,Jin:2017jjp}
in $\Omega_{\rm R}$ is not excluded in general.
In this case, the behavior of $B_4^\Omega$ near the CP
is modified from Eq.~(\ref{eq:B4L}) as~\cite{Jin:2017jjp}
\begin{align}
  B_4^\Omega( \lambda , LT )
  =& \big( b_4 + c ( \lambda-\lambda_{\rm c} ) (LT)^{1/\nu} \big)
   \big( 1 + d (LT)^{y_t-y_h} \big).
  \label{eq:B4fit6}
\end{align}
To investigate the effect of this possible mixing,
we try fits of $B_4^\Omega$ 
based on Eq.~(\ref{eq:B4fit6}).
We use the values of $B_4^\Omega$ at three $\lambda$
for the fits to increase the number of data points.
We find that the six parameter fits with Eq.~(\ref{eq:B4fit6})
with the fitting parameters $b_4$, $\lambda_{\rm c}$, $\nu$, $c$, $d$, $y_t-y_h$
are quite unstable, suggesting that
$\chi^2$ has many local minima.
The model space of Eq.~(\ref{eq:B4fit6}) would be too large against
the data.
As a next trial, we perform five parameter fits
with Eq.~(\ref{eq:B4fit6}) by fixing $y_t-y_h=-0.894$.
In this case,
  we find that $\chi^2$ still has many local minima, 
  and $\chi^2/{\rm dof}$ becomes larger compared with the four parameter fit.
  It is also found that the value of $d$ is consistent with zero
  within the error for all trials with the variation of $\lambda$ values.
This suggests that the mixing of the energy-like observable
in $\Omega_{\rm R}$ is negligible around the CP in the heavy-quark region.

\begin{table}
  \centering
  \caption{
  Location of the critical point $(\beta_{\rm c},\kappa_{\rm c})$ for various $N_{\rm f}$.
  For $\lambda_{\rm c}$, the first parentheses are for statistical errors and the second parentheses are for systematic errors from the fit as discussed in~Sec.~\ref{sec:B4}. The errors for $\beta_{\rm c}$ and $\kappa_{\rm c}$ include the systematic errors.
  }
  \label{table:Nf}
  \begin{tabular}{c|ccc}
   \hline
    $N_{\rm f}$ & $\beta_{\rm c}$ & $\kappa_{\rm c}$ & $\lambda_{\rm c}$
   \\ \hline
   $1$ & 5.68446(22) & 0.0714(5) & 0.00498(14)(2) \\
   $2$ & 5.68453(22) & 0.0602(4) & 0.00503(14)(2) \\
   $3$ & 5.68456(21) & 0.0544(4) & 0.00505(14)(2) \\
   \hline
 \end{tabular}
\end{table}

\subsection{$N_{\rm f}$ dependence}

In Table~\ref{table:Nf}, we summarize our final results
for the location of the CP, $(\beta_{\rm c},\kappa_{\rm c})$.
In the table, we also show the results for $N_{\rm f}=1$ and $3$. 
We note that the $N_{\rm f}$ dependence of the HPE is trivial
at the LO in the sense that $N_{\rm f}$ enters the action
Eq.~(\ref{eq:g+LO}) at this order
only through the combination $\lambda = 64 N_{\rm c} N_{\rm f}\kappa^4$
after the replacement $\beta\to\beta^*$.
Therefore, $\lambda_{\rm c}$ does not depend on $N_{\rm f}$.
At the LO, this allows us to obtain the value of $\kappa_{\rm c}$ for various 
$N_{\rm f}$ from the value of $\kappa_{\rm c}$ at $N_{\rm f}=2$~\cite{Saito:2013vja}.
Because such a simple scaling is no longer applicable at the NLO,
we made individual numerical analyses at $N_{\rm f}=1$ and $3$.
From Table~\ref{table:Nf}, we find that the results of $\lambda_{\rm c}$ are almost insensitive to $N_{\rm f}$.
This means that the NLO effects on $\lambda_{\rm c}$ are small.

\section{Distribution function of $\Omega_{\rm R}$}
\label{sec:p(O)}

In this section, we study the scaling behavior 
of the distribution function $p(\Omega_{\rm R})$ 
to further investigate the consistency with the $Z(2)$ universality class around the CP.

\subsection{Scaling of distribution function}
\label{sec:fssV}

\begin{figure}
  \centering
  \includegraphics[width=0.48\textwidth]{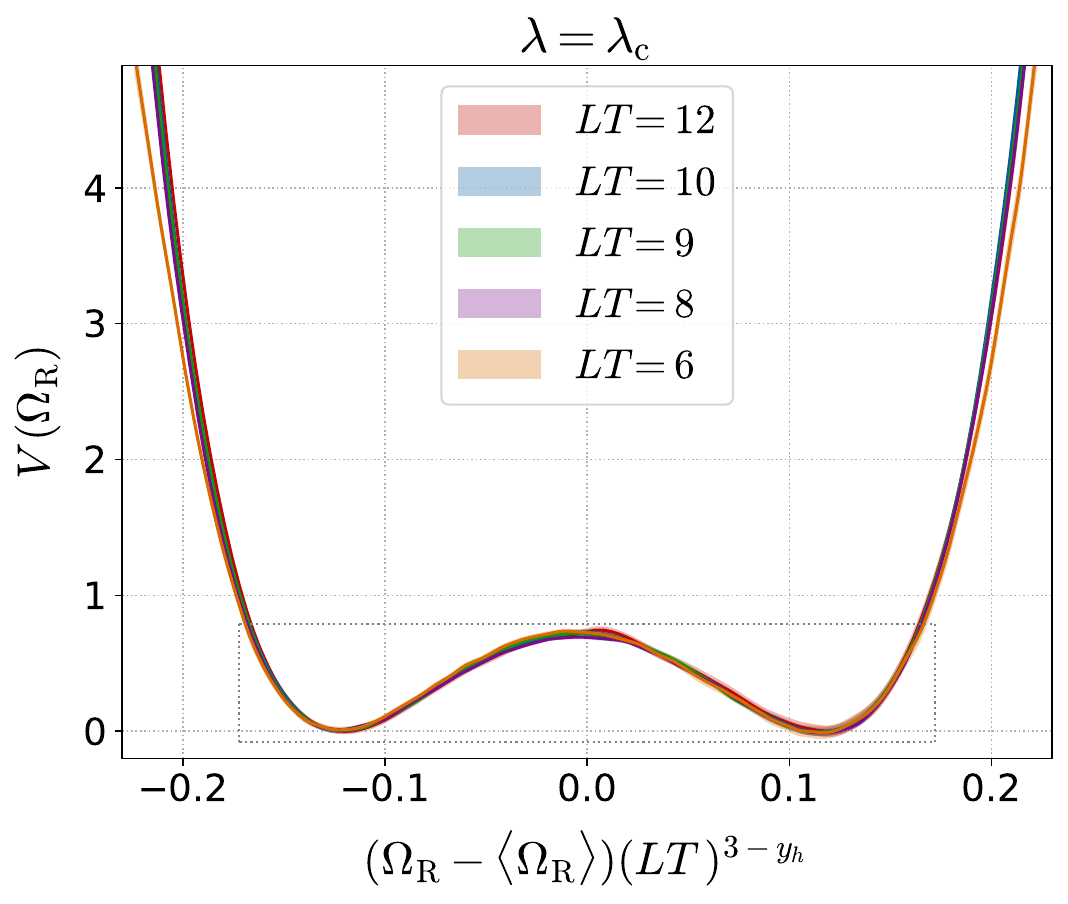}
  \includegraphics[width=0.48\textwidth]{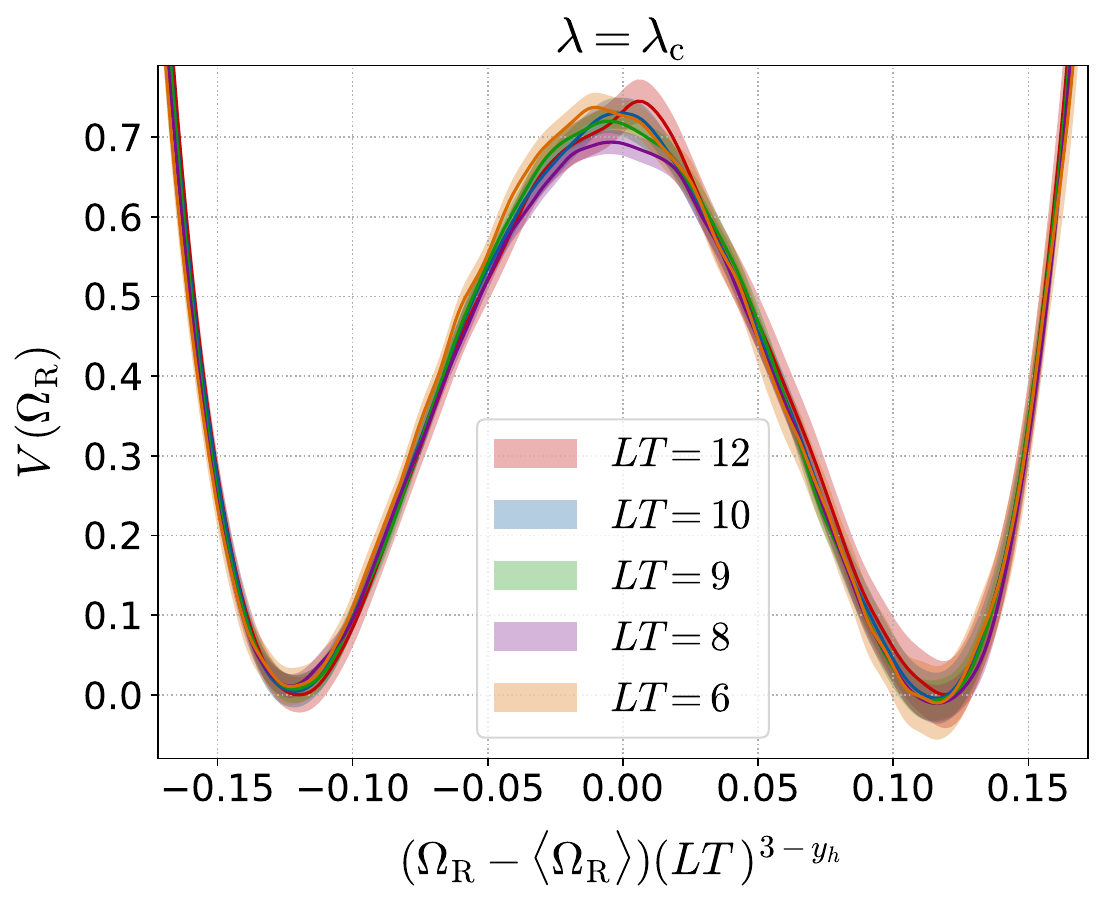}
  \caption{
    Effective potential $V(\Omega_{\rm R})= -\ln p(\Omega_{\rm R})$.
    Bottom panel is an enlargement of the region enclosed by a dotted rectangle in the top panel.
  }
\label{fig:V}
\end{figure}

Let us first focus on the $LT$ dependence of $p(\Omega_{\rm R})$ at the CP.
In the following, instead of $p(\Omega_{\rm R})$ itself,
we study the effective potential defined from $p(\Omega_{\rm R})$:
\begin{align}
  V(\Omega_{\rm R};\lambda,LT) = -\ln p(\Omega_{\rm R})_{\lambda,LT},
\end{align}
as this quantity is more convenient in comparing the results
at different $LT$~\cite{Saito:2011fs,Saito:2013vja}.
From Eq.~(\ref{eq:p_m-CP}), the $LT$ dependence of $V(\Omega_{\rm R},\lambda,LT)$ at the CP will be described
by a single function $\tilde{V}(x)$ as
\begin{align}
  V(\Omega_{\rm R};\lambda_{\rm c},LT)
  = \tilde{V}\Big(\big(\Omega_{\rm R}-\langle\Omega_{\rm R}\rangle \big) (LT)^{3-y_h}\Big),
  \label{eq:GO=tildeG}
\end{align}
up to an additive constant, 
where $\langle\Omega_{\rm R}\rangle$ is
subtracted from $\Omega_{\rm R}$ to adjust the center of the distribution.

To see if the scaling behavior of Eq.~(\ref{eq:GO=tildeG}) is satisfied,
we show in Fig.~\ref{fig:V} the effective potential 
$V(\Omega_{\rm R};\lambda_{\rm c},LT)$ at the CP obtained at five values of $LT$, 
as a function of $\big(\Omega_{\rm R}-\langle\Omega_{\rm R}\rangle \big) (LT)^{3-y_h}$, 
where we set $3-y_h=0.518$.
For the figure, we adjust the arbitrary constant term of 
$V(\Omega_{\rm R};\lambda_{\rm c},LT)$ 
such that $V(\Omega^{(1)};\lambda_{\rm c},LT)+V(\Omega^{(2)};\lambda_{\rm c},LT)=0$,
  where $\Omega^{(1)}$ and $\Omega^{(2)}~(>\Omega^{(1)})$ are the values of $\Omega_{\rm R}$
  at the two local minima of $V(\Omega_{\rm R};\lambda_{\rm c},LT)$.
No further adjustments are made in the figure.
The error bands do not include the uncertainty of the additive constant.
The lower panel is an enlargement of the region indicated by the dotted rectangle
in the upper panel.
  From Fig.~\ref{fig:V}, we find that the numerical results for
  $LT=8$--12 agree almost completely within the errors
  with the scaling relation Eq.~(\ref{eq:GO=tildeG}).
This result nicely supports the FSS in the $Z(2)$ universality class at the CP.
From the upper panel of Fig.~\ref{fig:V}, we note that
the effective potential for $LT=6$ shows a clear deviation from the results for larger volumes 
at $\Omega_{\rm R} \ll \Omega^{(1)}$ and $\Omega_{\rm R} \gg \Omega^{(2)}$, 
while it agrees well with them in the range $\Omega^{(1)} \lesssim \Omega_{\rm R} \lesssim \Omega^{(2)}$.
This suggests that the deviation from the $Z(2)$ FSS by lattices with small $LT$, discussed in Sec.~\ref{sec:B4}, is due to that in the tails of the distribution $p(\Omega_{\rm R})$ for small $LT$.

\subsection{Gap between the two minima}
\label{sec:gap}

Using Eq.~(\ref{eq:p_m}), the argument of Sec.~\ref{sec:fssV}  on the effective potential can be extended
away from the CP along the transition line.
In this subsection, we study
the gap between the two local minima of
$V(\Omega_{\rm R};\lambda_{\rm c},LT)$,
\begin{align}
  \Delta \Omega = \Omega^{(2)} - \Omega^{(1)}.
  \label{eq:DeltaO}
\end{align}
According to Eq.~(\ref{eq:minimum}),
this quantity should behave around the CP as
\begin{align}
  \Delta \Omega(\lambda,LT)
  = (LT)^{y_h-3} \Delta\tilde\Omega \big( (\lambda-\lambda_{\rm c}) (LT)^{1/\nu} \big)
  \label{eq:DeltaOmega}
\end{align}
provided that $p(\Omega_{\rm R})$ obeys the FSS.

\begin{figure}
  \centering
  \includegraphics[width=0.48\textwidth]{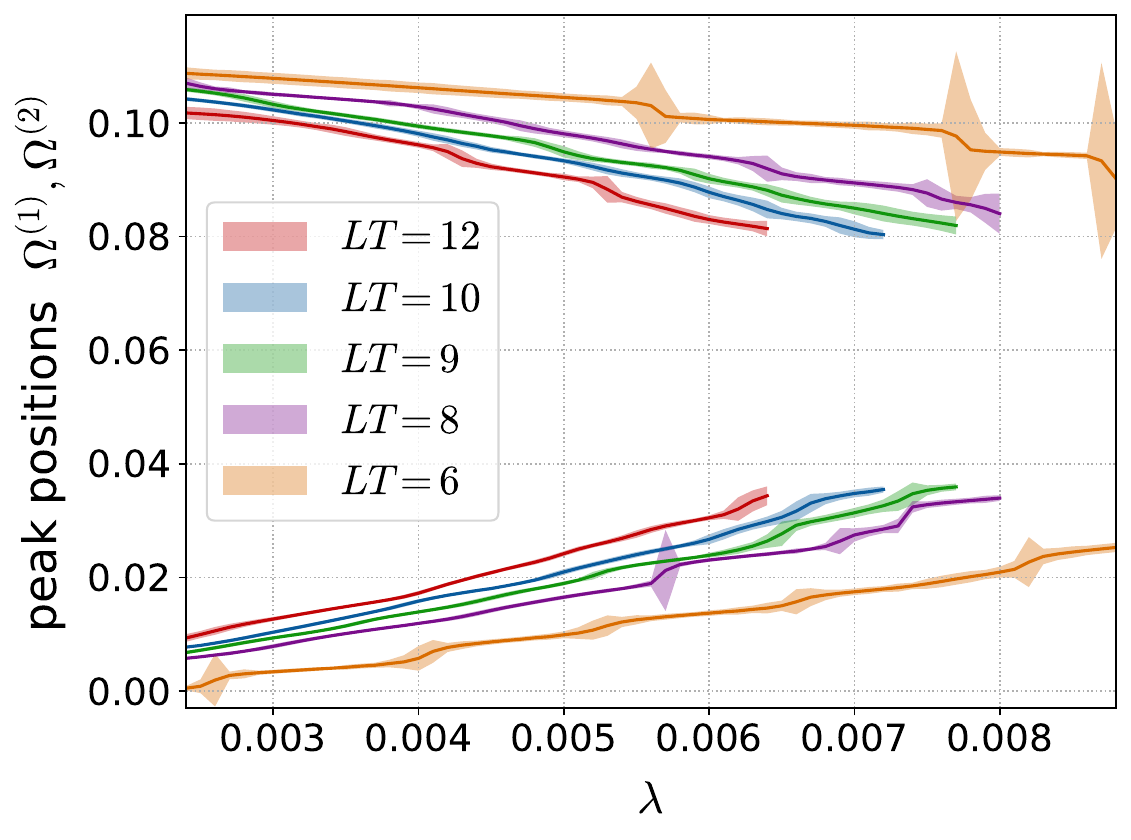}
  \caption{
    Positions of peaks of the distribution function $p(\Omega_{\rm R})$ measured on the transition line.
  }
\label{fig:Delta}
\end{figure}

In Fig.~\ref{fig:Delta}, we show the $\lambda$ dependence of
$\Omega^{(1)}$ and $\Omega^{(2)}$. 
As seen from Fig.~\ref{fig:distr}, a clear two peak structure of $p(\Omega_{\rm R})$ disappears when $\lambda$ exceeds some value depending on $LT$. Even before the disappearance of the two peaks, 
identification of local maxima of $p(\Omega_{\rm R})$ becomes unstable by statistical fluctuations.
In Fig.~\ref{fig:Delta}, we thus truncate the plots for $\Omega^{(1)}$ and $\Omega^{(2)}$ at finite $\lambda$.
The shaded areas in the figure represent statistical errors estimated by the jackknife method, 
for which we repeat the analysis of $\Omega^{(1,2)}$ for $p(\Omega_{\rm R})$
obtained in each jackknife sample with the smearing width of 
$\Delta_{\Omega_{\rm R}}=0.002$.
\footnote{ 
We see that the errors in Fig.~\ref{fig:Delta} become occasionally large. 
We find that this is due to statistical oscillations in the shape of $p(\Omega_{\rm R})$ around the peak:
Though the oscillations are within the statistical errors, the peak position in each jackknife sample can jump discontinuously
when oscillation appears just at the peak position as we vary $\lambda$.
This makes the resulting jackknife error large there.
From this observation, we think that these large errors are overestimated.
}
As shown in Appendix~\ref{sec:smear},
$\Delta_{\Omega_{\rm R}}$ dependence of these results
is well suppressed at this $\Delta_{\Omega_{\rm R}}$.

\begin{figure}
  \centering
  \includegraphics[width=0.48\textwidth]{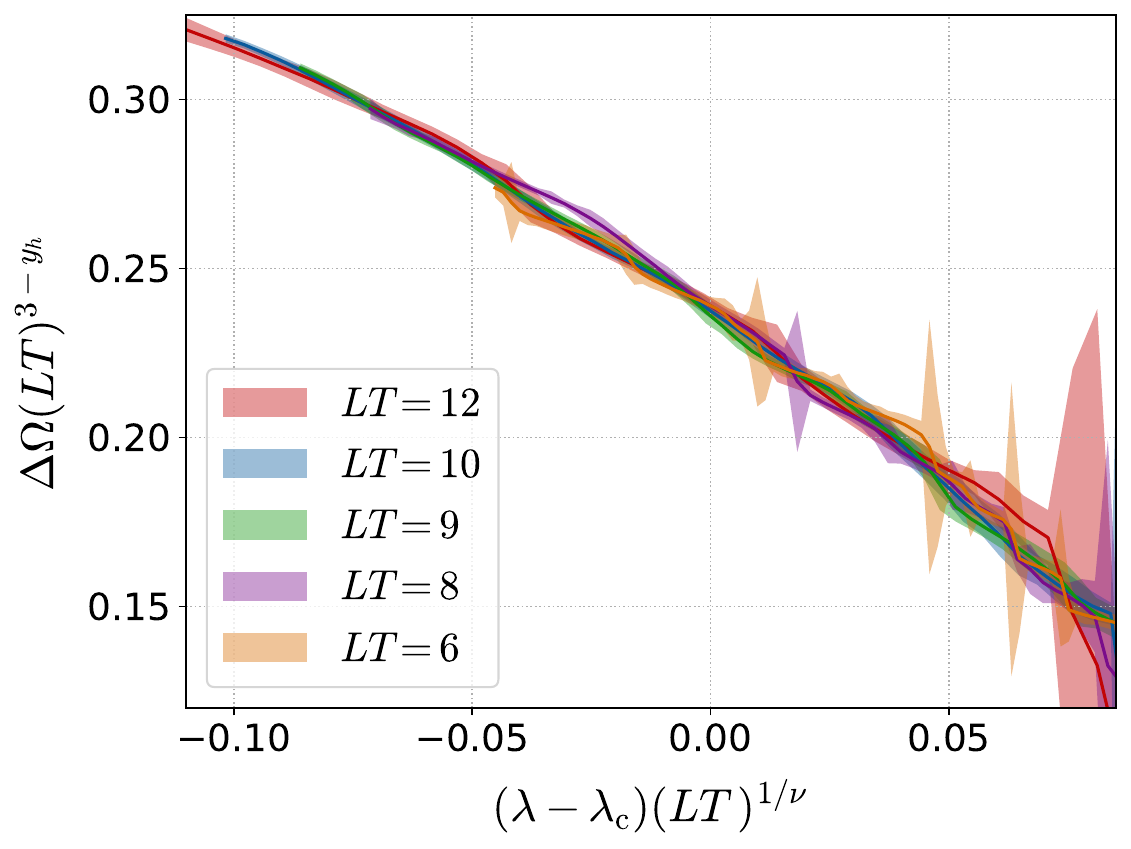}
  \caption{
     Scaling of the gap $\Delta\Omega$ around $T_{\rm c}$.
  }
\label{fig:Delta-scale}
\end{figure}

From Fig.~\ref{fig:Delta} we extract $\Delta\Omega$
as a function of $\lambda$.
In Fig.~\ref{fig:Delta-scale}, we show $\Delta\Omega$ for five different volumes.
To see the FSS,
the vertical and horizontal axes are adjusted according to
Eq.~(\ref{eq:DeltaOmega}), where
the $Z(2)$ values $3-y_h=0.518$ and $\nu=0.630$,
and $\lambda_{\rm c}=0.00503$ determined in the previous section
are used.
The figure shows that, 
for a wide range of $\lambda-\lambda_{\rm c}$ and $LT$, 
the results of $\Delta\Omega$ obtained on different volumes are on top of each other within the errors. 
This supports the FSS of $p(\Omega_{\rm R})$ around the
peak positions over a wide range of $LT$ and $\lambda$.

It is interesting to note that the scaling behavior of $\Delta\Omega$ 
is observed even at $LT=6$, although the FSS of $B_4^\Omega$ is
violated already at $LT=8$.
As discussed in the previous subsection, 
we may understand this when the violation of the FSS for $B_4^\Omega$ is due to the violation in the tails of the distribution function $p(\Omega_{\rm R})$.
As seen in Fig.~\ref{fig:V}, 
$V(\Omega_{\rm R})$ at various volumes agrees well for
$\Omega^{(1)} \lesssim \Omega \lesssim \Omega^{(2)}$ even for small values of $LT$.
As the higher-order cumulants are sensitive to the whole structure of the distribution,
$B_4^\Omega$ will be more sensitive to the violation of FSS at the tails of the distribution.
On the other hand, $\Delta\Omega$ is insensitive to them by definition. 
We also note that the statistical error of $\Delta\Omega$ is naturally small because it is defined by the peaks of the distribution.
Therefore, $\Delta\Omega$ is useful in seeing the FSS around the CP.

\begin{figure}
  \centering
  \includegraphics[width=0.48\textwidth]{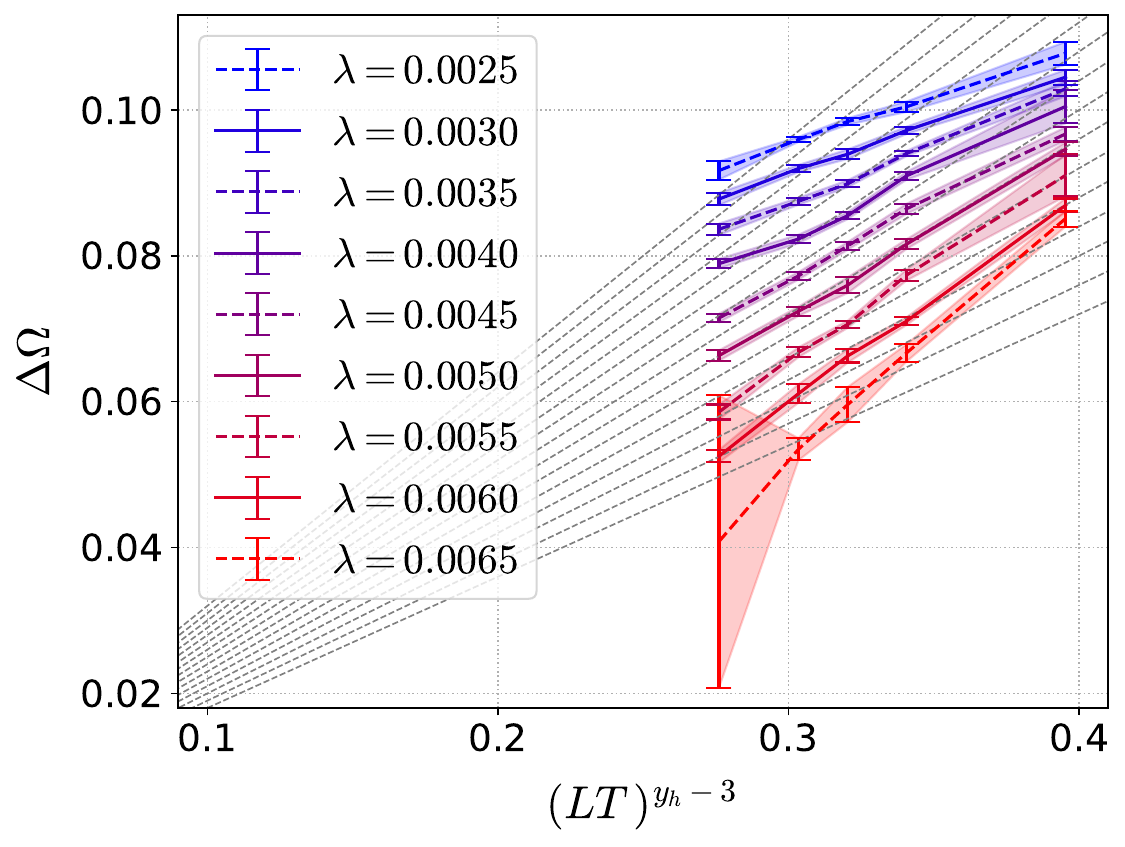}
  \caption{
    Gap $\Delta\Omega$ around $T_{\rm c}$ as a function of $(LT)^{y_h-3}$. The $Z(2)$ value $3-y_h=0.518$ is assumed.
  }
\label{fig:Delta-Ns}
\end{figure}

From Eq.~(\ref{eq:DeltaOmega}), $\Delta\Omega$ should behave linearly
as a function of $(LT)^{y_h-3}$ at the CP $\lambda=\lambda_{\rm c}$.
In Fig.~\ref{fig:Delta-Ns}, we show $\Delta\Omega$ on the transition line at various values of $\lambda$, 
as a function of $(LT)^{y_h-3}$. 
In the same figure, the dashed lines show linear functions
$\Delta\Omega = k (LT)^{y_h-3}$ for various values of $k$.
Figure~\ref{fig:Delta-Ns} suggests that the linear behavior is realized
at $\lambda\simeq0.005$, which is consistent with our estimation $\lambda_{\rm c}=0.00503(14)(2)$
from the analysis of $B_4^\Omega$.

\subsection{Discussions}

Let us comment on the relation of the present results with
those given in Refs.~\cite{Saito:2011fs,Saito:2013vja,Ejiri:2019csa}.
In these studies, 
the CP is defined as the point at which the two peak structure of 
$p(\Omega_{\rm R})$ disappears. 
On lattice with finite $LT$, this leads to $\lambda$ which is larger than our value of $\lambda_{\rm c}$ in the $L\to\infty$ limit.
In fact, in Ref.~\cite{Saito:2011fs} the location of the CP
is estimated as $\kappa_{\rm c}=0.0658(3)(^{+4}_{-11})$ for $N_{\rm f}=2$ and $N_t=4$, which is about 10\% larger than that given in Table~\ref{table:Nf}.
Values of $\lambda_{\rm c}$ ($\kappa_{\rm c}$) which are smaller than Ref.~\cite{Ejiri:2019csa} for each $N_t$ 
are also reported by 
a recent study of the CP in the heavy quark region on fine lattices ($N_t=6$--$10$) 
using the Binder cumulant method~\cite{Cuteri:2020yke}.
Though the difference may be removed in the $L\to\infty$ limit, a careful extrapolation will be required.
Because the FSS is clearly identified in this study, we think that the extrapolation to the large volume limit is stably performed with the present analysis.

We also note that the latent heat at the deconfinement transition
in the $SU(3)$ Yang-Mills theory ($\kappa=0$) has been measured
in Ref.~\cite{Shirogane:2020muc} recently.
It was found that the latent heat becomes larger
with increasing the spatial volume.
This may be attributed to a remnant of the FSS around 
the $Z(2)$ CP, like $\Delta\Omega$ studied in the present study.

\section{Conclusions}
\label{sec:conclusion}

In this paper, 
we studied the distribution function of the Polyakov loop 
and its cumulants around the CP in the heavy quark region of QCD.
Large volume simulations up to the aspect ratio $N_s/N_t=LT=12$ have been carried out
to see the finite-size scaling, while the lattice spacing is fixed to $N_t=4$.
We have performed the measurement of observables using the hopping parameter expansion for
the quark determinant; the measurement has been performed
at the next-to-leading order of the hopping parameter expansion by the multipoint reweighting method evaluated
on the gauge configurations generated for the action at the leading order.
We found that this analysis is quite effective in reducing statistical
errors by avoiding the overlapping problem of the reweighting method,
while the numerical cost hardly changes from the pure Yang-Mills simulations.
The convergence of the hopping parameter expansion at the next-to-leading order at the critical point is also verified by the
comparison with the leading order result.

Using the data on $p(\Omega_{\rm R})$ thus obtained,
we have performed the Binder cumulant analysis for determining
the location of the critical point and evaluating the critical exponent.
We found that the critical exponent $\nu$
and the value of the Polyakov-loop Binder cumulant $B_4^\Omega$ at the critical point is consistent with the
$Z(2)$ universality class when $LT\ge9$ data are used for the analysis.
On the other hand, statistically-significant deviation from
the $Z(2)$ scaling is observed when the data at
$LT=8$ is included, which suggests that this spatial volume
is not large enough to apply the finite-size scaling.

The scaling behavior near the critical point is further studied using
the structure of the distribution function of the real part of the Polyakov loop, $p(\Omega_{\rm R})$.
We found that the structure of $p(\Omega_{\rm R})$
for various $LT$ obeys the finite-size scaling well especially near the peaks
of $p(\Omega_{\rm R})$.
We have also proposed the use of the gap $\Delta\Omega$ between the peaks of
$p(\Omega_{\rm R})$ for the finite-size scaling analysis.
We showed that the $\lambda$ and $LT$ dependence of 
$\Delta\Omega$ is in good agreement with the $Z(2)$ scaling
over a wide range of $\lambda$ and $LT$.
On the other hand, the deviation of $p(\Omega_{\rm R})$
around the tails of the distribution is observed on small
lattices, which would give rise to the violation of the finite-size scaling of
$B_4^\Omega$ in small volumes.

\vspace{5mm}
\noindent\textbf{Acknowledgments}

The authors thank Frithjof Karsch, Macoto Kikuchi, Yoshifumi Nakamura, and the members of WHOT-QCD Collaboration for useful discussions.
This work was supported by in part JSPS KAKENHI (Grant Nos.~JP17K05442, JP18K03607, JP19H05598, 
JP19K03819, JP19H05146,  JP21K03550), 
the Uchida Energy Science Promotion Foundation, 
the HPCI System Research project (Project ID: hp170208, hp190036, hp200089, hp210039), and Joint Usage/Research Center for Interdisciplinary Large-scale Information Infrastructures in Japan (JHPCN) (Project ID: jh190003, jh190035, jh190063, jh200049).
This research used computational resources of 
OCTPUS of the large-scale computation program at the Cybermedia Center, Osaka University,
and ITO of the JHPCN Start-Up Projects at the Research Institute for Information Technology, Kyushu University.

\appendix

\section{Cumulants}
\label{sec:cumulant}

Let us consider a probability distribution function $p(x)$
of a stochastic variable $x$.
Since $p(x)$ represents probability,
it satisfies the normalization condition $\int \! dx \, p(x)=1$.

The $m$th-order moment $\langle x^m \rangle$ of $p(x)$ is defined by
\begin{align}
  \langle x^m \rangle = \int \! dx\,  x^m p(x).
\end{align}
Using the moment generating function
\begin{align}
G(\theta) = \int \! dx \, e^{x\theta} p(x) = \langle e^{x\theta} \rangle,
\label{eq:G(theta)}
\end{align}
the moments are also given by
\begin{align}
\langle x^m \rangle = \partial_\theta^m G(\theta)|_{\theta=0},
\label{eq:moment}
\end{align}
with $\partial_\theta=\partial/\partial\theta$.

The cumulants are defined from the cumulant generating function
\begin{align}
  K(\theta) = \ln G(\theta) = \ln \langle e^{x\theta} \rangle
  \label{eq:K}
\end{align}
as
\begin{align}
\langle x^m \rangle_{\rm c} = \partial_\theta^m K(\theta)|_{\theta=0}.
\label{eq:cumulant}
\end{align}
From Eqs.~(\ref{eq:moment}) and (\ref{eq:cumulant}),
one easily finds that the cumulants are related to the moments as
\begin{align}
  \langle x \rangle_{\rm c} 
  &= \langle x \rangle,
  \\
  \langle x^2 \rangle_{\rm c} 
  &= \langle x^2 \rangle - \langle x \rangle^2 
  = \langle \delta x^2 \rangle ,
  \\
  \langle x^3 \rangle_{\rm c} &= \langle \delta x^3 \rangle ,
  \label{eq:<m3>c=}
  \\
  \langle x^4 \rangle_{\rm c} 
  &= \langle \delta x^4 \rangle - 3 \langle \delta x^2 \rangle^2,
  \label{eq:<m4>c=}
\end{align}
and etc. with $\delta x = x - \langle x \rangle$.
The cumulants are useful in representing properties of $p(x)$
than the moments for various purposes.
In particular, in statistical mechanics 
the cumulants of an extensive variable are extensive variables;
see, for example, Ref.~\cite{Asakawa:2015ybt}.

\section{Effect of smearing width for distribution function}
\label{sec:smear}

\begin{figure*}[t]
  \centering
  \includegraphics[width=0.32\textwidth]{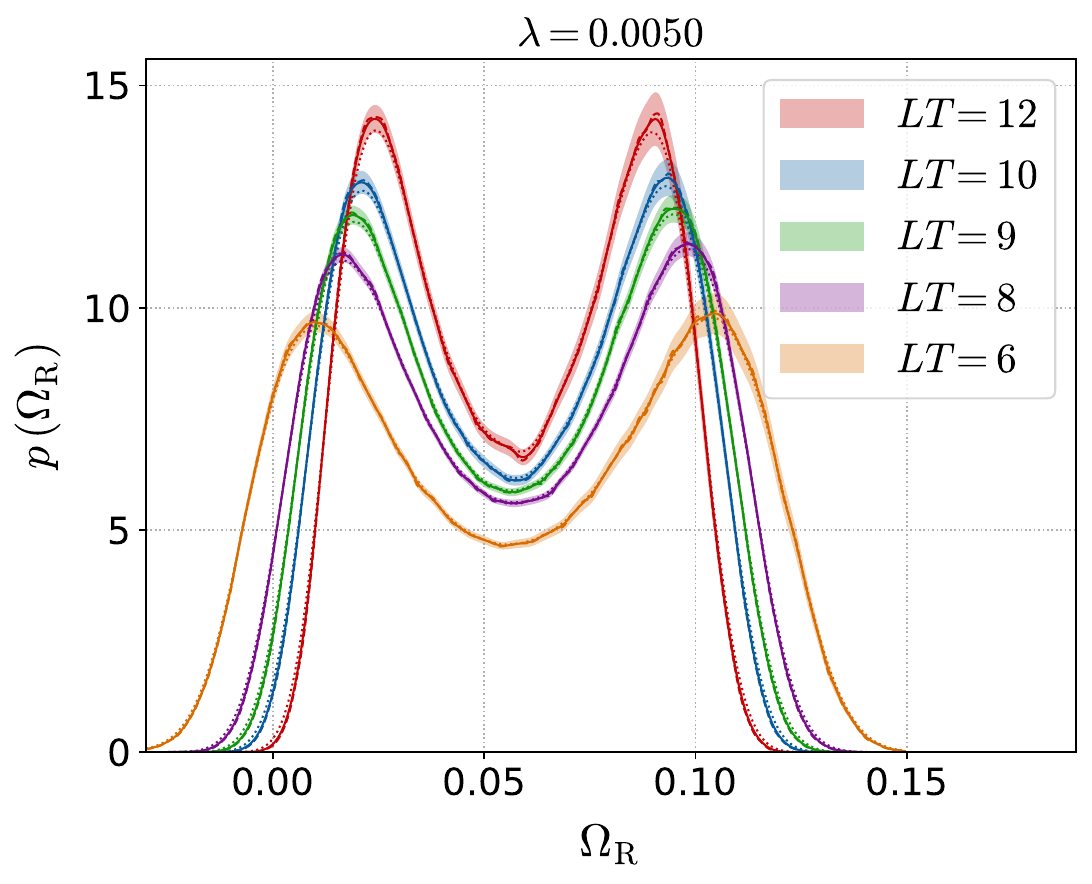}
  \includegraphics[width=0.32\textwidth]{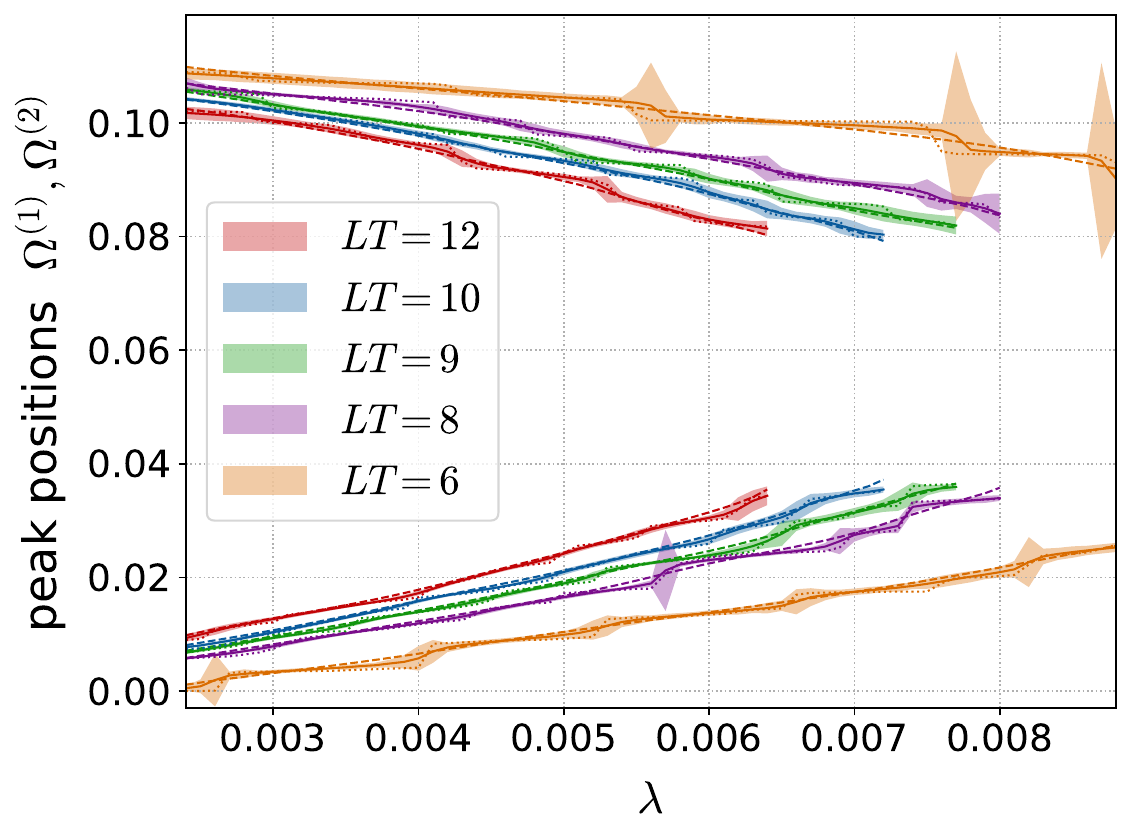}
  \includegraphics[width=0.32\textwidth]{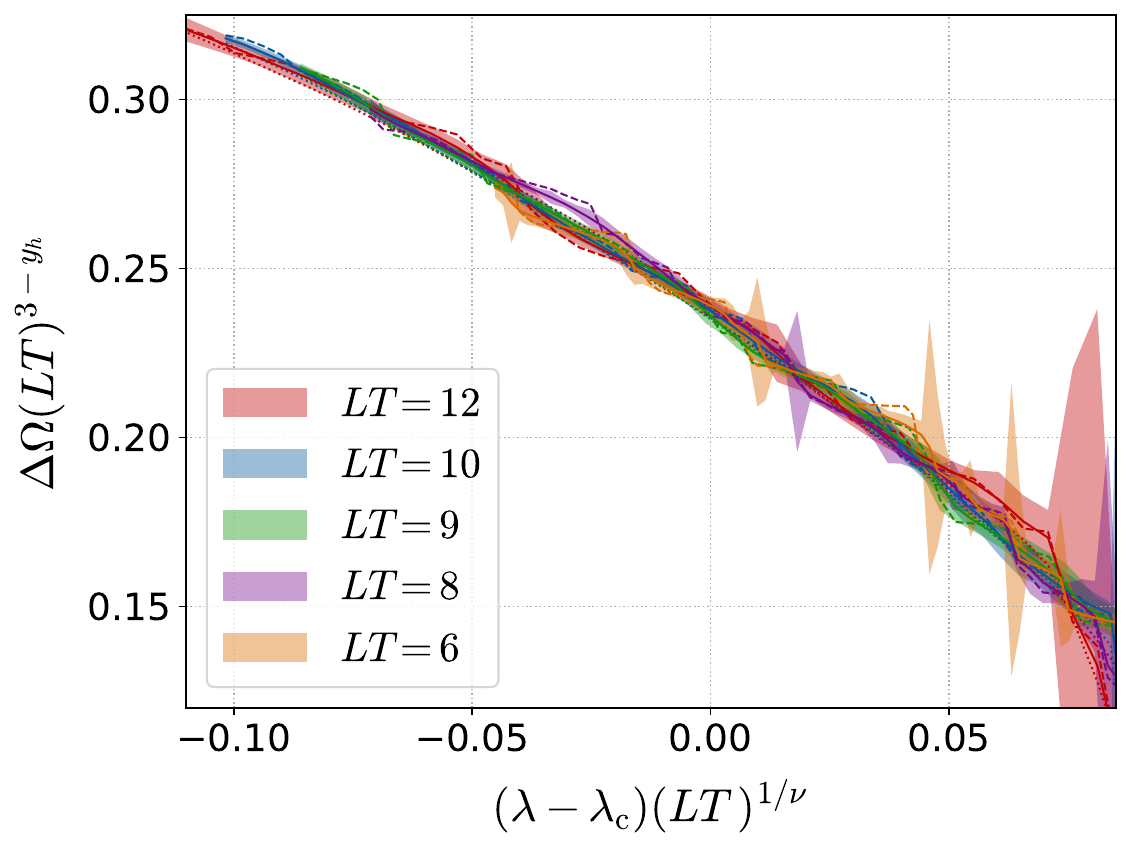}
  \caption{
    $\Delta_{\Omega_{\rm R}}$ dependence of the results given in
    Secs.~\ref{sec:Binder} and \ref{sec:p(O)}.
    The left, middle and right panels correspond to the middle panel
    of Fig.~\ref{fig:distr}, Fig.~\ref{fig:Delta} and
    Fig.~\ref{fig:Delta-scale}, respectively.
    The dashed and dotted lines in each panel shows the results
    with $\Delta_{\Omega_{\rm R}}=0.001$ and $0.004$, respectively,
    while the solid lines with $\Delta_{\Omega_{\rm R}}=0.002$ are
    the same as the original figures.
  }
  \label{fig:smear}
\end{figure*}

In Secs.~\ref{sec:Binder} and \ref{sec:p(O)}, we calculate the 
distribution function $p(\Omega_{\rm R})$ with smearing
the delta function in Eq.~(\ref{eq:distr})
by the normalized Gauss function with the width $\Delta_{\Omega_{\rm R}}$. 
With the statistics of this study, we choose $\Delta_{\Omega_{\rm R}}=0.002$ from an examination of the statistical error and resolution of $p(\Omega_{\rm R})$.

In this Appendix,
we examine the $\Delta_{\Omega_{\rm R}}$ dependence of the numerical results,
picking up the middle panel of Fig.~\ref{fig:distr}, Fig.~\ref{fig:Delta}
and Fig.~\ref{fig:Delta-scale} as representative results.
In Fig.~\ref{fig:smear}, we compare these results with those obtained with 
$\Delta_{\Omega_{\rm R}}=0.001$ and $0.004$.
In Fig.~\ref{fig:smear}, 
the results with $\Delta_{\Omega_{\rm R}}=0.001$ and $0.004$
are shown by the dashed and dotted lines, while the solid lines show
the results with $\Delta_{\Omega_{\rm R}}=0.002$.
From this figure, we find that $\Delta_{\Omega_{\rm R}}$ dependence is well small around $\Delta_{\Omega_{\rm R}}=0.002$ and does not affect our conclusions.

\section{LO and NLO hopping parameter expansion}
\label{sec:HPEapp}

In this appendix we derive Eqs.~(\ref{eq:LO}) and (\ref{eq:NLO}).
Throughout this Appendix we assume general value for $N_t$.

\begin{figure}[tb]
  \centering
  \includegraphics[width=0.17\textwidth]{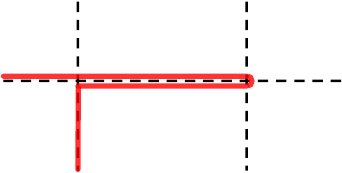}
  \caption{
    ``Appendix'' structure of trajectories.
  }
\label{fig:appendix}
\end{figure}

As in Eq.~(\ref{eq:HPE}),
the HPE of $\ln [ \det M(\kappa)]$ is given by ${\rm Tr}[B^n]$.
Since the matrix $B$ has nonzero contributions
only between neighboring lattice sites,
${\rm Tr}[B^n]$ are graphically represented by
the closed trajectories with $n$ links~\cite{Rothe:1992nt}.
However, trajectories including ``appendices''
shown in Fig.~\ref{fig:appendix} do not contribute
to the HPE because for such a path the product of the matrix in Dirac
space vanishes at the tip of the appendix
as $(1-\gamma^\mu)(1+\gamma^\mu)=0$~\cite{Rothe:1992nt}.
With this exception, all possible closed trajectories composed of $n$
links contribute to the HPE at the order $\kappa^n$.
In the following, we calculate their contributions by classifying 
the trajectories by the shape.

Let us start from the plaquette, \textit{i.e.} $1\times1$ rectangle,
which gives a lowest-order contribution to Eq.~(\ref{eq:HPE}) at
the order $\kappa^4$.
The plaquette operator $\hat{P}$ in Eq.~(\ref{eq:Plaq})
is defined in such a way that $\langle \hat{P} \rangle=1$
in the weak coupling limit with $U_\mu(x)=1$.
To satisfy this condition Eq.~(\ref{eq:Plaq}) has a coefficient
$1/(N_{\rm c} M_{\rm plaq})=1/18$, where $M_{\rm plaq}$ is the number of
different plaquettes per lattice site;
$M_{\rm plaq}={}_4{\rm C}_2=6$, which is 
the number of combinations of axes $(\mu,\nu)$
at which the plaquettes are located.

The contribution from all plaquettes to Eq.~(\ref{eq:HPE}) is calculated to be
\begin{align}
  -2 N_{\rm c} M_{\rm plaq} D_{\rm plaq} N_{\rm site} \kappa^4 \hat{P}.
  \label{eq:HPEplaq}
\end{align}
Here, $D_{\rm plaq}$ is the coefficient from the trace in the Dirac space
\begin{align}
  D_{\rm plaq} = {\rm tr_D} [ (1-\gamma^\mu)(1-\gamma^\nu)(1+\gamma^\mu)(1+\gamma^\mu)] = -8,
\end{align}
where ${\rm tr_D}$ means the trace over the Dirac indices.
The factor $2$ in Eq.~(\ref{eq:HPEplaq}) comes from two directions for
each trajectory, which have to be distinguished in the HPE.
The factor $1/n$ in Eq.~(\ref{eq:HPE}) is canceled
by the number of four starting points of a trajectory;
this cancellation occurs for all trajectories.

Next, let us consider the Wilson loops of length $6$ without
windings along the temporal direction.
At this order there are three types of trajectories;
$1\times2$ rectangle, chair, crown, which are shown in Fig.~\ref{fig:6step}.
We define the operators, $\hat{W}_{\rm rect}$,
$\hat{W}_{\rm chair}$, $\hat{W}_{\rm crown}$, corresponding to these
trajectories as 
\begin{widetext}
 \begin{align}
 \hat{W}_{\rm rect}
 =& \frac1{N_{\rm c} M_{\rm rect}N_{\rm site}}
 \sum_{x}  \sum_{\mu \ne \nu}
 {\rm Re} \ {\rm tr_C} \left[ U_{x,\mu} U_{x+\hat\mu,\mu} U_{x+2\cdot\hat{\mu},\nu}
U^{\dagger}_{x+\hat\mu+\hat\nu,\mu} U^{\dagger}_{x+\hat\nu,\nu} U^{\dagger}_{x,\nu} \right],
 \label{eq:rect}
 \\
 \hat{W}_{\rm chair}
 =& \frac1{N_{\rm c} M_{\rm chair}N_{\rm site}}
 \sum_{x} \sum_{\mu,\nu<\rho,\nu\ne\mu\ne\rho} \sum_{s,t=\pm1}
    {\rm Re} \ {\rm tr_C}  \left[ U_{x,s\nu} U_{x+s\hat\nu,\mu} U_{x+\hat\mu,s\nu}^\dagger
       U_{x+\hat\mu,t\rho} U_{x+t\hat\rho,\mu}^\dagger U_{x,t\rho}^\dagger \right],
 \label{eq:chair}
 \\
 \hat{W}_{\rm crown}
 =& \frac1{N_{\rm c} M_{\rm crown}N_{\rm site}}
 \sum_{x} \sum_{\mu<\nu<\rho} \sum_{s,t=\pm1}
     {\rm Re} \ {\rm tr_C}  \left[ U_{x,\mu} U_{x+\hat\mu,s\nu} U_{x+\hat\mu+s\hat\nu,t\rho}
       U_{x+s\hat\nu+t\hat\rho,\mu}^\dagger U_{x+t\hat\rho,s\nu}^\dagger U_{x,t\rho}^\dagger \right],
 \label{eq:crown}
\end{align}
\end{widetext}
with $U_{x,-\mu}=U^\dagger_{x-\hat{\mu},\mu}$ and
$U_{x,-\mu}^\dagger=U_{x-\hat{\mu},\mu}$ for $\mu>0$.
Eqs.~(\ref{eq:rect})--(\ref{eq:crown}) are defined so that
$\langle\hat{W}_{\rm rect}\rangle=\langle\hat{W}_{\rm chair}\rangle
=\langle\hat{W}_{\rm crown}\rangle=1$
in the weak coupling limit again;
to satisfy this condition, the operators are divided by
$M_{\rm rect}=12$, $M_{\rm chair}=48$, and $M_{\rm crown}=16$, respectively, 
corresponding to the number of trajectories per lattice site.

The contribution of these trajectories to the HPE of $\ln\det M(\kappa)$
is given by
\begin{align}
  -2 N_{\rm c} \sum_s M_s D_s N_{\rm site} \kappa^6 \hat{W}_s,
  \label{eq:HPEdim6}
\end{align}
with $s=$rect, chair, crown.
$D_s$ is the coefficient from the trace in the Dirac space.
For the $1\times2$ rectangle we have
\begin{align}
  D_{\rm rect} =& \,{\rm tr_D}
  \big[ (1-\gamma_\mu) (1-\gamma_\mu) (1-\gamma_\nu)
  \nonumber \\
  &\times (1+\gamma_\mu)(1+\gamma_\mu)(1+\gamma_\nu) \big]
  = -32,
\end{align}
and similar manipulations lead to $D_{\rm chair}=D_{\rm crown}=-16$.

Next we consider the Polyakov-loop type operators having a winding
along the temporal direction.
The lowest-order contribution among them is the Polyakov loop,
\textit{i.e.} the straight lines of length $N_t$.
To calculate its contribution,
one needs to pay a special attention to the fact that
there is only one independent Polyakov loop for each 
{\it spatial coordinate} on one time slice, not for each lattice site.
Therefore, their contributions to the HPE is given by
\begin{align}
  2 N_{\rm c} D_{\rm pol} N_s^3 \kappa^{N_t} \hat\Omega_{\rm R},
\end{align}
where the factor $-1$ is to be applied because of the anti-periodic boundary
condition of the quark determinant.
The real part of $\hat\Omega$ has to be taken after multiplying the factor 2
because two directions of a trajectory are independently taken into account.
The factor from the Dirac trace for the Polyakov loop
is calculated to be
\begin{align}
  D_{\rm pol} = {\rm tr_D} \big[ ( 1-\gamma_4)^{N_t} \big] = 2^{N_t+1}.
\end{align}

Finally, we consider the contribution of the bent Polyakov loops
shown in Fig.~\ref{fig:bent}, whose explicit definitions are given by
\begin{widetext}
  \begin{eqnarray}
\hat\Omega_1 &=& \frac{1}{N_{\rm c} M_{\rm bent} N_{\rm site}}\sum_x
\sum_{i=1}^3 \sum_{s=\pm1}
{\rm tr_C}  \Big[ \;
U_{x,s i} U_{x+s\hat{i},4} U_{x+\hat4,s i}^{\dagger} U_{x+\hat4,4}
U_{x+2\cdot\hat4,4} \cdots U_{x+(N_t-1)\cdot\hat4,4} 
\Big] ,
\label{eq:bendpl1}
\\
\hat\Omega_2 &=& \frac{1}{N_{\rm c} M_{\rm bent} N_{\rm site}} \sum_{x}
\sum_{i=1}^3 \sum_{s=\pm1}
{\rm tr_C}  \Big[ \;
U_{x,s i} U_{x+s\hat{i},4} U_{x+s\hat{i}+\hat4,4}
U_{x+2\cdot\hat4,s i}^{\dagger} U_{x+2\cdot\hat4,4} \cdots U_{x+(N_t-1)\cdot\hat4,4} 
\Big] ,
\label{eq:bendpl2}
  \end{eqnarray}
\end{widetext}
and so on. 
The factor $M_{\rm bent}=6$ is needed to make
$\langle \hat\Omega_n \rangle=1$ in the weak coupling limit.
  From the definition of $\hat\Omega_n$ we have
  $\hat\Omega_n=\hat\Omega_{N_t-n}$.
  Also, when $N_t$ is even 
  $\hat\Omega_{N_t/2}$ counts each trajectory twice,
  and thus its contribution to the HPE should be divided by $2$.
  Bearing these facts in mind,
  the contribution from $\hat\Omega_n$ to the HPE of $\ln\det M(\kappa)$
  is given by
\begin{align}
  & 2 N_{\rm c} M_{\rm bent} D_{\rm bent} N_{\rm site}
  \kappa^{N_t+2}
  \nonumber \\
  &\qquad \times
  \Big( \sum_{n=1}^{N_t/2-1} {\rm Re}\,\hat\Omega_n
  + \frac12{\rm Re}\,\hat\Omega_{N_t/2} \Big)
\end{align}
where the last term ${\rm Re}\,\hat\Omega_{N_t/2}$ should be omitted
for odd $N_t$.
The contribution of the Dirac trace is calculated to be $D_{\rm bent}=2^{N_t+1}$.

Accumulating these results gives Eqs.~(\ref{eq:LO}) and (\ref{eq:NLO}).

\section{Comparison of LO and NLO results}
\label{sec:LO}

\begin{figure}[t]
  \centering
  \includegraphics[width=0.48\textwidth]{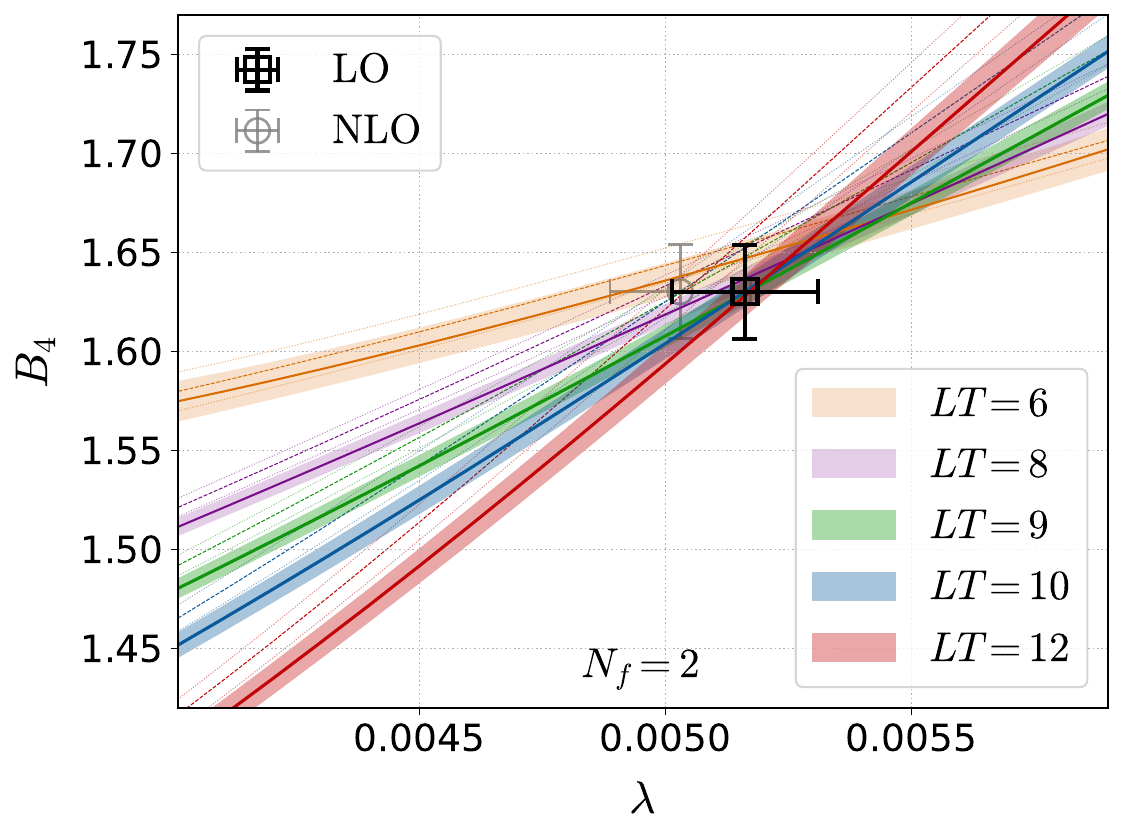}
  \caption{
    Binder cumulant $B_4$ as a function of $\lambda$
    calculated at the LO. The results at the NLO are
    also plotted by the thin dashed lines.
    The square with errors shows the fit result with three largest volumes.
    The NLO result on the same fit result is also shown by the
    thin circle.
  }
\label{fig:B4-LO}
\end{figure}

In this appendix, to see the convergence of the HPE at the NLO,
we repeat the analyses in Sec.~\ref{sec:B4} at the LO and compare
its results with those at the NLO.
In Fig.~\ref{fig:B4-LO}, we show the 
Binder cumulant $B_4^\Omega$ obtained at the LO along the transition line.
In this figure, the NLO results of Fig.~\ref{fig:B4} are also shown by the thin dashed lines.
We find that, though the difference between the LO and the NLO results grows as $\lambda$ becomes larger, the
deviation is within a few percent level around the crossing point $\lambda_{\rm c}$.

In Fig.~\ref{fig:B4-LO},  we also show the result of the four parameter fit with Eq.~(\ref{eq:B4fit4}) 
using data at the LO on $LT=12$, 10 and 9 lattices by the black square.
The same result at the NLO is shown by gray circle for comparison. 
We find that the LO result of the fit is consistent with the NLO result within statistical errors:
The values of $b_4=1.630(24)$ and $\nu=0.620(47)$ at the LO are hardly changed from the corresponding NLO values given in Table~\ref{table:B4Ns}.
Though the central value of $\lambda_{\rm c}=0.00516(15)$ at the LO is about $2.6\%$ larger than the NLO value, it is consistent with the NLO result within errors, suggesting that the truncation error of the HPE at the NLO is well suppressed in these quantities.

The success of the $B_4^\Omega$ fit together with the consistency of the fit results with the $Z(2)$ values suggests that the $Z(2)$ scaling is realized also with the LO action when the system volume is sufficiently large.
This is reasonable since the scaling properties near the CP are insensitive to detailed structures of the theory.

\bibliographystyle{apsrev4-1}
\bibliography{refs.bib}

\end{document}